\documentclass[aps,prb,twocolumn,superscriptaddress,draft,showpacs,intlimits,amsmath,amssymb,floats]{revtex4}
\usepackage{bm}
\usepackage[final]{graphicx}
\usepackage{epsfig}
\usepackage{color}
\definecolor{DarkBlue}{rgb}{0.1,0.1,0.5}
\definecolor{Red}{rgb}{0.9,0.0,0.1}
\definecolor{Green}{rgb}{0.0,0.99,0.0}

\begin{document}

\title{Evidence for weak electronic correlations in Fe-pnictides}

\author{W.~L.~Yang}
\affiliation{Advanced Light Source, Lawrence Berkeley National Laboratory, Berkeley, CA 94720}
\author{A.~P.~Sorini}
\affiliation{Stanford Institute for Materials and Energy Sciences, SLAC National Accelerator Laboratory, 2575 Sand Hill Road, Menlo Park, CA 94025, USA.}
\author{C-C.~Chen}
\affiliation{Stanford Institute for Materials and Energy Sciences, SLAC National Accelerator Laboratory, 2575 Sand Hill Road, Menlo Park, CA 94025, USA.}
\affiliation{Department of Physics, Stanford University, Stanford, California 94305, USA}
\author{B.~Moritz}
\affiliation{Stanford Institute for Materials and Energy Sciences, SLAC National Accelerator Laboratory, 2575 Sand Hill Road, Menlo Park, CA 94025, USA.}
\author{W.-S.~Lee}
\affiliation{Stanford Institute for Materials and Energy Sciences, SLAC National Accelerator Laboratory, 2575 Sand Hill Road, Menlo Park, CA 94025, USA.}
\author{F.~Vernay}
\affiliation{Paul Scherrer Institut, Condensed Matter Theory Group, Villigen PSI, Switzerland}
\author{P.~Olalde-Velasco}
\affiliation{Advanced Light Source, Lawrence Berkeley National Laboratory, Berkeley, CA 94720}
\affiliation{Instituto de Ciencias Nucleares, UNAM, 04510 Mexico DF, Mexico}
\author{J.~D.~Denlinger}
\affiliation{Advanced Light Source, Lawrence Berkeley National Laboratory, Berkeley, CA 94720}
\author{B.~Delley}
\affiliation{Paul Scherrer Institut, Condensed Matter Theory Group, Villigen PSI, Switzerland}
\author{J.-H.~Chu}
\affiliation{Stanford Institute for Materials and Energy Sciences, SLAC National Accelerator Laboratory, 2575 Sand Hill Road, Menlo Park, CA 94025, USA.}
\affiliation{Department of Applied Physics, Stanford University, Stanford, California 94305, USA}
\affiliation{Geballe Laboratory for Advanced Materials, Stanford University, Stanford, CA 94305, USA.}
\author{J. G. Analytis} 
\affiliation{Stanford Institute for Materials and Energy Sciences, SLAC National Accelerator Laboratory, 2575 Sand Hill Road, Menlo Park, CA 94025, USA.}
\affiliation{Department of Applied Physics, Stanford University, Stanford, California 94305, USA}
\affiliation{Geballe Laboratory for Advanced Materials, Stanford University, Stanford, CA 94305, USA.}
\author{I.~R.~Fisher}
\affiliation{Stanford Institute for Materials and Energy Sciences, SLAC National Accelerator Laboratory, 2575 Sand Hill Road, Menlo Park, CA 94025, USA.}
\affiliation{Department of Applied Physics, Stanford University, Stanford, California 94305, USA}
\affiliation{Geballe Laboratory for Advanced Materials, Stanford University, Stanford, CA 94305, USA.}
\author{Z.~A.~Ren}
\affiliation{National Lab for Superconductivity, Institute of Physics, Chinese Academy of Sciences, Beijing, P. R. China.}
\author{J.~Yang}
\affiliation{National Lab for Superconductivity, Institute of Physics, Chinese Academy of Sciences, Beijing, P. R. China.}
\author{W.~Lu}
\affiliation{National Lab for Superconductivity, Institute of Physics, Chinese Academy of Sciences, Beijing, P. R. China.}
\author{Z.~X.~Zhao}
\affiliation{National Lab for Superconductivity, Institute of Physics, Chinese Academy of Sciences, Beijing, P. R. China.}
\author{J.~van den Brink}
\affiliation{Stanford Institute for Materials and Energy Sciences, SLAC National Accelerator Laboratory, 2575 Sand Hill Road, Menlo Park, CA 94025, USA.}
\affiliation{Institute Lorentz for Theoretical Physics, Leiden University, P.O. Box 9506, 2300 RA Leiden, The Netherlands}
\author{Z.~Hussain}
\affiliation{Advanced Light Source, Lawrence Berkeley National Laboratory, Berkeley, CA 94720}
\author{Z.-X.~Shen}
\affiliation{Stanford Institute for Materials and Energy Sciences, SLAC National Accelerator Laboratory, 2575 Sand Hill Road, Menlo Park, CA 94025, USA.}
\affiliation{Department of Applied Physics, Stanford University, Stanford, California 94305, USA}
\affiliation{Geballe Laboratory for Advanced Materials, Stanford University, Stanford, CA 94305, USA.}
\affiliation{Department of Physics, Stanford University, Stanford, California 94305, USA}
\author{T.~P.~Devereaux}
\affiliation{Stanford Institute for Materials and Energy Sciences, SLAC National Accelerator Laboratory, 2575 Sand Hill Road, Menlo Park, CA 94025, USA.}
\affiliation{Geballe Laboratory for Advanced Materials, Stanford University, Stanford, CA 94305, USA.}

\date{\today}

\begin{abstract}
Using x-ray absorption and resonant inelastic x-ray scattering, charge dynamics at and near the Fe $L$ edges is investigated in Fe pnictide materials, and contrasted to that measured in other Fe compounds. It is shown that the XAS and RIXS spectra for 122 and 1111 Fe pnictides are each qualitatively similar to Fe metal. 
Cluster diagonalization, multiplet, and density-functional calculations show that Coulomb correlations are much smaller than in the cuprates, highlighting the role of Fe metallicity and strong covalency in these materials. 
Best agreement with experiment is obtained using Hubbard parameters
 $U\lesssim 2$eV and $J\approx 0.8$eV.
\end{abstract}


\maketitle

\section{Introduction}

In the new and rapidly developing field of Fe-pnictide superconductivity, the question of what constitutes the basic ingredients for high transition temperatures remains largely unanswered. Parallels have been drawn to the cuprate high-temperature superconductors, which contain partially filled $d$-electron spins that in the parent phase are aligned antiferromagnetically like the pnictides, and high-temperature superconductivity emerges when magnetism can be suppressed. Common to many ideas is that superconductivity itself may be emergent from the two competing phases, driven by an underlying quantum critical point. 

A key question that needs to be addressed to understand this framework is whether or not the Fe pnictides are strongly correlated like the cuprates. Since density functional theory (DFT) calculations have indicated that the electron-phonon interaction is too weak to account for high transition temperatures,\cite{eph1,eph2,eph3} the strength of the Coulomb correlations would give some account for the pairing strength possible in an electronic-based pairing mechanism.\cite{Kivelson} Recent renormalization group flow\cite{DHLee} and random-phase approximation (RPA) calculations of effective tight-binding models fitted to DFT bands\cite{Graser1,Graser2} indicate that several pairing instabilities, such as sign-changing $s-$wave pairing and $d-$type pairing, all have nearly the same energy, which depends subtly on Coulomb parameters $U$ and magnetic exchange $J$. Therefore pinning down these numbers would greatly focus the discussion of the physics of the Fe pnictides. 

Theoretically, the situation is complicated. Traditional DFT methods, which can be extremely accurate in uncorrelated materials, can account for the correct atomic structure but yield large sublattice magnetic moments that have not been observed in experiments.\cite{Mazin} This overestimation of the magnetic moment is exactly opposite to the situation in the cuprates, where DFT underestimates the moment, and implies that Coulombic effects are a small part of the story for the pnictides. However it does not provide an explanation as to why the moments are so much larger than those found in experiments\cite{Giovannetti08}. 

Recently, theoretical treatments using combinations of DFT and dynamical mean field theory (DMFT) have yielded opposite conclusions.\cite{Haule,Craco,Georges,Anisimov1,Anisimov2,Anisimov3} One set of calculations yield Hubbard parameters $U\sim 4$ eV, giving a Kondo-like peak near the Fermi level and a well separated lower Hubbard band, and argue that these materials are on the verge of a Mott transition.\cite{Haule,Craco} Another set however gives $U\sim 1$ eV.\cite{Georges,Anisimov1,Anisimov2,Anisimov3} Angle-resolved photoemission (ARPES) studies have shown strong density of states (DOS) near the Fermi level relative to the pnictogen valence band.\cite{Lu} The Fe conduction band states, with a bandwidth $W~4$ eV, as well as overall band dispersions can be well matched to density functional calculations with a mass renormalization of 2. This is in direct contrast with the cuprates, with valence band spectral weight spread out over much larger energy scales, and conduction bands having much larger band renormalizations, smaller bandwidths and DOS. Thus it would appear from ARPES that these materials are categorically different than the cuprates.

X-ray measurements have been crucial in uncovering the physics of the cuprates, identifying multi-particle states, such as the Zhang Rice singlet, as signatures of strong $d-$level Coulomb interactions, as well as excitons and satellites due to strong core-hole interactions.\cite{Degroot} These strong satellites and spread out spectral weights have not been observed in recent x-ray absorption (XAS) measurements at the oxygen $K$-edge in 1111, setting an upper bound on the effective Coulomb parameter $U\sim 1$ eV.\cite{Kroll,Moewes1,Moewes2} Absorption and emission studies on the Fe $L_{2,3}$ edges also are in agreement with weak electronic correlations, and can be simply matched to the unoccupied $d$ DOS determined from DFT calculations. However calculations pertinent to the XAS process, where core holes are created, and the emission process, where photons are emitted from the valence states in the presence of a core hole, have not been carried out. This is crucially needed in order to understand the true role of Coulomb interaction in these materials.

In this paper we present a comprehensive study of XAS measurements and resonant inelastic x-ray scattering (RIXS) at the Fe $L_{2,3}$ edges in a variety of Fe-based materials, including the superconducting 1111 Fe pnictide SmO$_{0.85}$FeAs and the undoped 122 pnictides BaFe$_2$As$_2$ and LaFe$_2$P$_2$. It is shown that the XAS spectra of Fe pnictides look qualitatively, and in some cases quantitatively, similar to Fe metal and show no features resembling the multiple peak structures seen in Fe insulators, such as hematite ($\alpha$-Fe$_2$O$_3$) and other iron oxides. A resonance study of x-ray emission across the $L_2$ and $L_3$ edges demonstrates that the RIXS spectra is dominated by fluorescence, with no observance of discernable excitonic or satellite peaks. 

In addition, we present calculations using three separate models which specifically include and account for the role of the core hole in x-ray absorption and emission processes. These calculations are performed using quantum cluster, multiplet, and DFT-based methods, and highlight the roles of Fe metallicity, FeAs covalency, and local Coulomb and Hund's couplings. DFT calculations using {\tt FEFF}\cite{FEFF} give quantitative agreement with XAS measurements and align absorption peaks to Fe d-DOS 
%(and not As) 
above the Fermi level, demonstrating the minor role of core hole interactions. Cluster calculations of XAS support the role of strong Fe-As hybridization involving As states below the Fermi level, setting an upper bound of $U\sim 2$eV. This indicates that the FeAs materials are weakly correlated and that the physics is governed largely by Fe metallicity.

The outline of paper is as follows. In Sec. \ref{Expt}, XAS and RIXS measurements on superconducting SmO$_{0.85}$FeAs, non-superconducting BaFe$_2$As$_2$, LaFe$_2$P$_2$, $\alpha$-Fe$_2$O$_3$, and Fe metal are presented, comparing and contrasting qualitative behaviors across these compounds, while in Sec. \ref{Theory} calculations are presented for XAS and XES at the Fe $L_{2,3}$ edges. Secs. \ref{cluster} and \ref{multiplet} present calculations for $L$-edge XAS in Fe clusters to highlight the expected role of strong Coulomb correlations, and it is shown that spectral features related to the strong correlations that are not seen in experiments of Sec. \ref{Expt} can be used to set upper limits on Hubbard parameters. Moreover, DFT-based {\tt FEFF} calculations, which include multiple scattering and core hole effects, are presented in Sec. \ref{FEFF}, and are shown to provide excellent agreement with the measured XAS spectra. Finally, Sec. \ref{summary} summarizes our findings and states our conclusions.

\section{XAS and RIXS Measurements}
\label{Expt}
The SmO$_{0.85}$FeAs samples with superconducting transition temperature (T$_c$) of 55K, so far the highest T$_c$ in the family of iron arsenides, were prepared by a high-pressure synthesis method\cite{ZhiAn1,ZhiAn2}. Sample quality was checked by x-ray powder diffraction and T$_c$ was confirmed by both transport and magnetic measurements\cite{ZhiAn1,ZhiAn2}. We have also measured F-doped samples with the same T$_c$, as well as the non-superconducting parent compounds SmOFeAs, but found no obvious difference in the spectra. BaFe$_2$As$_2$ and LaFe$_2$P$_2$ single crystals were prepared by the flux method \cite{Fishercrystals1,Fishercrystals2,fish}. Data shown here were collected at room temperature with incident beam 45 degrees to sample surfaces. We noticed serious surface oxidization effects for the iron pnictides, and to avoid this surface oxidization problem all the data were collected on {\it in-situ} cleaved sample surfaces.

\begin{figure}[h!]
\includegraphics[width=2.75in]{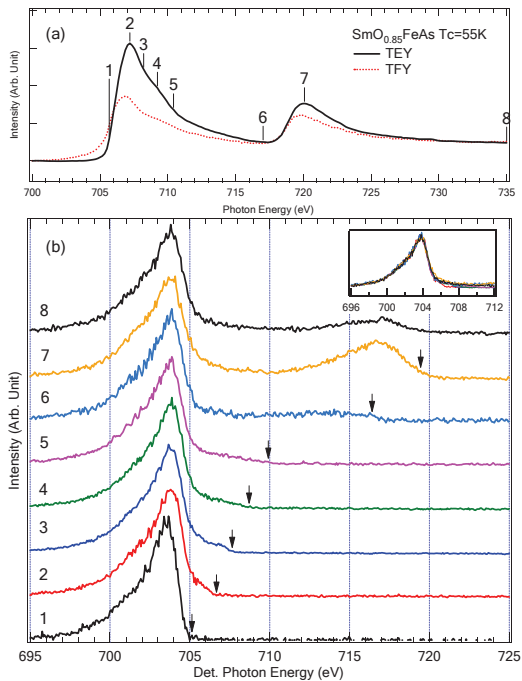}
\caption{\label{Fig:1}
(a) X-ray absorption spectra of the 55K T$_c$ SmO$_{0.85}$FeAs. TEY and TFY intensity of the Fe $L_{2,3}$ edges is plotted as a function of incident photon energy. The difference between TEY and TFY data is mainly from the self-absorption effect in TFY. (b) RIXS spectra of SmO$_{0.85}$FeAs collected with excitation energy across the Fe $L_{2,3}$ absorption peaks. The number on the left stands for the excitation energy corresponding to the number marked in (a), the value of which is marked with the arrows on the spectra. Inset shows the prominent Fe $L_3$ emission peak collected with excitation energy above the Fe $L_3$ absorption edge (no.2 to 8), they all overlap nicely with the nonresonant spectrum (no.8). Note that RIXS spectra were normalized to the Fe $L_3$ emission peak for emphasizing the similar lineshape.}
\end{figure}

\begin{figure}[h!]
\includegraphics[width=2.75in]{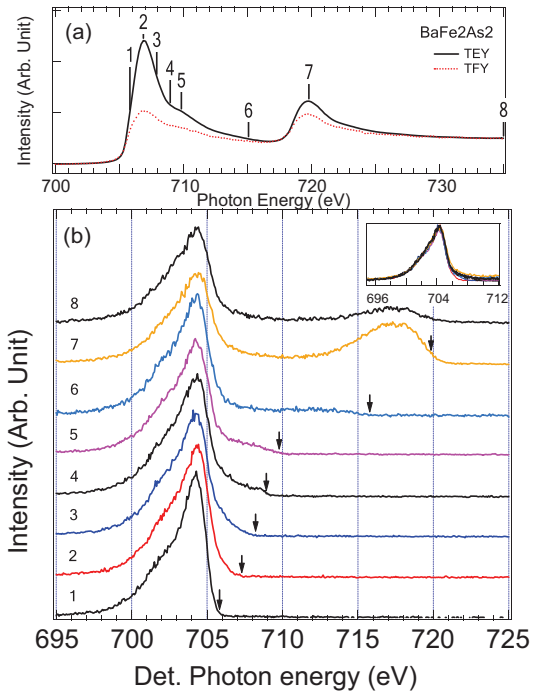}
\caption{\label{Fig:2}
(a) Fe $L_{2,3}$ XAS spectra of a BaFe$_2$As$_2$ single crystal. (b) RIXS spectra of BaFe$_2$As$_2$ collected with excitation energy labeled and marked in (a). Inset shows that all the Fe $L_3$ emission peaks (scaled to the same intensity) collected with excitation energy above Fe $L_3$ absorption edge (no.2 to 8) overlap with the nonresonant spectrum (no.8).}
\end{figure}

XAS and RIXS measurements were performed at beamline 8.0 of the Advanced Light Source at Lawrence Berkeley National Laboratory. The undulator and spherical grating monochromator supply a linearly polarized photon beam with resolving power up to 6000. RIXS data were collected by a Rowland circle geometry spectrometer \cite{Callcott} perpendicular to the incident beam. The linear polarization of the incident beam is parallel to the scattering plane. XAS spectra were collected by measuring sample current (TEY) and fluorescent yield (TFY). All XAS spectra have been normalized to the beam flux measured by a clean gold mesh. The resolution is better than 0.2eV for XAS measurements. For the X-ray emission measurements, the incident beam resolution is about 0.9eV and the spectrometer resolution is about 0.7eV.

The Fe $L_{2,3}$ absorption structure of iron pnictides are shown on top of Figs.~1-3. According to dipole selection rules, iron is a 3d element displaying $L_{2,3}$ absorption features from $2p^6$$3d^6$ to $2p^5$$3d^7$ transitions. The spin-orbit interaction splits the 2p states into 2p$_{1/2}$ and 2p$_{3/2}$, leading to two well separated peaks. The intensity ratio of the two peaks is largely defined by the high-spin or low-spin ground states related to the crystal field \cite{Thole}. As the $2p$ core levels are featureless and narrow, $L_2$ and $L_3$ absorption peaks often provide detailed information on the electronic structure of the unoccupied 3d states. As shown in Figs.~1a-3a, all the iron pnictide samples, including the 55K T$_c$ SmO$_{0.85}$FeAs (Fig.1a), non-superconducting BaFe$_2$As$_2$ (Fig.2a) and LaFe$_2$P$_2$ (Fig.3a), exhibit only the two major peaks, $L_2$ at about 720eV and $L_3$ at about 707eV. There are weak shoulders around 709.5eV, but no peak splitting or intensity ratio change was observed. This result is consistent with that on other 1111 \cite{Kroll}and 122 compounds \cite{Moewes2}. As XAS has been demonstrated to be a powerful tool for probing the crystal field and electronic interactions for 3d metals \cite{grootprb, Thole, Laan}; the non-splitting XAS structure indicates a weak crystal field effect \cite{crystalfield} that favors high-spin ground states.

Fig.1b shows the RIXS data of the superconducting SmO$_{0.85}$FeAs obtained at energies labeled in Fig.1a. The spectrum on top (No. 8) was collected with an incident photon energy of 735eV, which is far above the Fe $L_2$ and $L_3$ absorption edges, the so called nonresonant normal emission spectrum. Like all other 3d transition metals, this nonresonant spectrum exhibits two main fluorescent features at about 704eV and 717eV, resulting from the refill of the 2p$_{3/2}$ and 2p$_{1/2}$ holes respectively. The 2p$_{1/2}$ feature is very weak compared to the $L_2$ edge in the XAS spectrum, partially due to the Coster-Kronig decay process of the $2p_{1/2}$ holes to $2p_{3/2}$ \cite{CKdecay}.

The RIXS spectra collected with resonant energies also display the strong 704eV peak as seen in the nonresonant spectrum. With the excitation energy approaching the $L_3$ absorption edge (No.1), the 704eV peak evolves and stays at the same energy with all the excitations above the $L_3$ edge (No.2-7). This fluorescent feature does not track the excitation energy and overlap with the 704eV peak in the nonresonant spectrum (inset of Fig.1b). No energy loss feature, which is normally associated with various electron excitations and correlations, was displayed by the RIXS data.

RIXS of nonsuperconducting BaFe$_2$As$_2$ (Fig.2b) and LaFe$_2$P$_2$ (Fig.3b) share the same characterization as that of the 55K superconducting SmO$_0.85$FeAs, also in agreement with RIXS data reported on another 122 compound \cite{Moewes2}. Charge excitation features like Kondo peak and lower Hubbard peak are completely absent, and the RIXS data is dominated by a peak at 704eV with the only difference being the strength of the 701.5eV shoulder. Again, the prominent peaks collected at different resonant energies overlap nicely with the fluorescent peak in the nonresonant spectrum (insets of Fig.2b and 3b).

\begin{figure}
\includegraphics[width=2.75in]{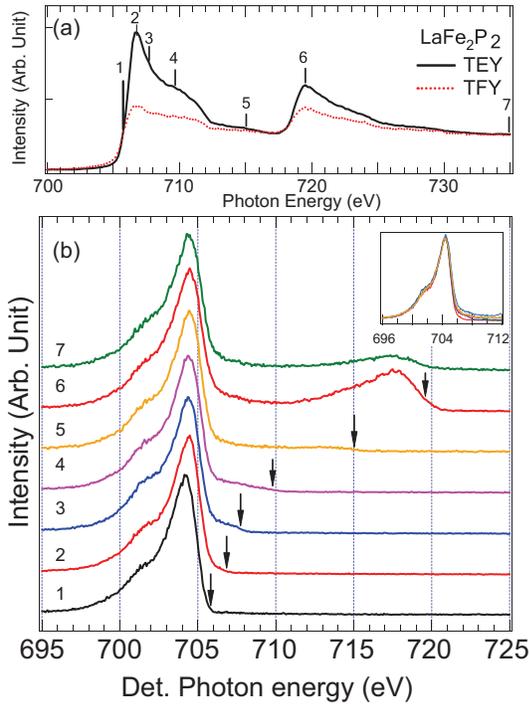}
\caption{\label{Fig:3}
(a) Fe $L_{2,3}$ XAS spectra of a LaFe$_2$P$_2$ single crystal. (b) RIXS spectra of BaFe$_2$P$_2$ collected with excitation energy labeled and marked in (a). Inset shows that all the Fe $L_3$ emission peaks (scaled to the same intensity) collected with excitation energy above Fe $L_3$ absorption edge (no.2 to 7) overlap with the nonresonant spectrum (no.7).}
\end{figure}

\begin{figure}
\includegraphics[width=2.75in]{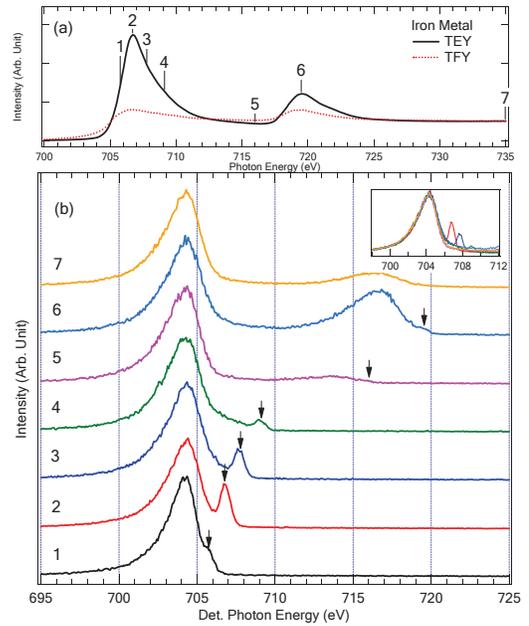}
\caption{\label{Fig:4}
(a) Fe $L_{2,3}$ XAS spectra of Fe metal. (b) RIXS spectra of Fe metal collected with excitation energy labeled and marked in (a). Inset shows that, with the elastic peak tracking excitation energy, all the Fe $L_3$ emission peaks collected with excitation energy above Fe $L_3$ absorption edge (no.2 to 7) overlap with the nonresonant spectrum (no.7), same as that of iron pnictides.}
\end{figure}

The absence of excitation induced energy-loss features in the RIXS data for all the iron pnictide samples indicates the weak correlation in this system. It is thus desirable to compare the iron pnictides with known metallic and insulating iron components, to reveal the importance of metallicity and to show that this result is not experimental resolution limited. In Fig.4, we show the XAS and RIXS data collected on pure iron metal. The XAS (Fig.4a) displays non-splitting $L_2$ and $L_3$ peaks \cite{Femetal1,Femetal2,Femetal3} with a slightly weaker shoulder compared to the iron pnictides. The RIXS data of Fe metal show more symmetric peaks without shoulders, as well as stronger elastic peaks tracking the excitation energies. But just like the iron pnictides, the RIXS lineshape is dominated by the peak at 704eV which overlaps with the fluorescence peak collected with off resonance excitation energy (inset of Fig.4b), and the iron metal resembles all the iron pnictides in the featureless RIXS data without excitation or correlation peaks.

On the contrary, the $\alpha$-Fe$_2$O$_3$ powder sample displays very different XAS and abundant features in RIXS. 
The XAS data (Fig.5a) shows strong splitting structure on both $L_2$ and $L_3$ absorption edges due to the interplay of crystal-field ($10Dq$=0.88eV) and 
electronic interactions \cite{Laan}. 
RIXS data (Fig.5b) show that the spectral appearance changes drastically with excitation energies, and obviously do not overlap with the nonresonant spectrum (No. 10). We plotted the energy loss features at different resonant energes in Fig.5c. Particular energy loss features, as indicated by the gray lines, were enhanced at particular resonant energies, leading to very different lineshape. These energy loss features are signatures of {\it dd}-excitations \cite{DudaFe2O3}, details on which is not the topic of this paper. With better resolution, our RIXS data revealed more excitation modes than that in previous publications, covering the whole range of the optic absorption bands \cite{Fe2O3optic}.

\begin{figure}
\includegraphics[width=2.75in]{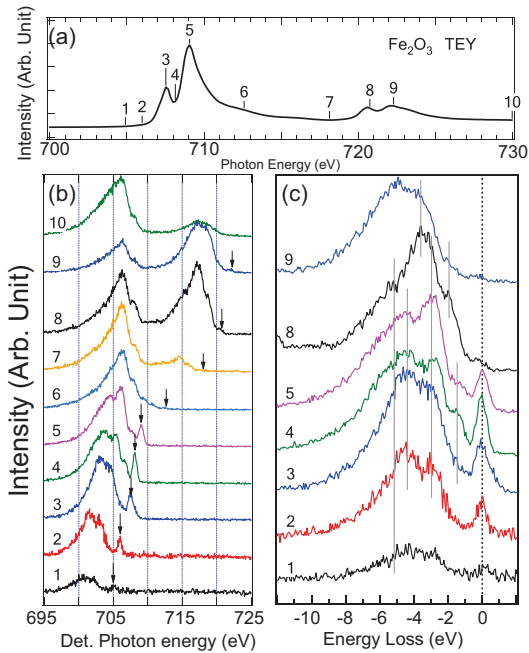}
\caption{\label{Fig:5}
(a) Fe $L_{2,3}$ XAS spectra of $\alpha$-Fe$_2$O$_3$ powder. (b) RIXS spectra of $\alpha$-Fe$_2$O$_3$ collected with excitation energy labeled and marked in (a). (c) Energy loss features corresponding to {\it dd}-excitations marked with gray lines.}
\end{figure}

\begin{figure}
\includegraphics[width=2.75in]{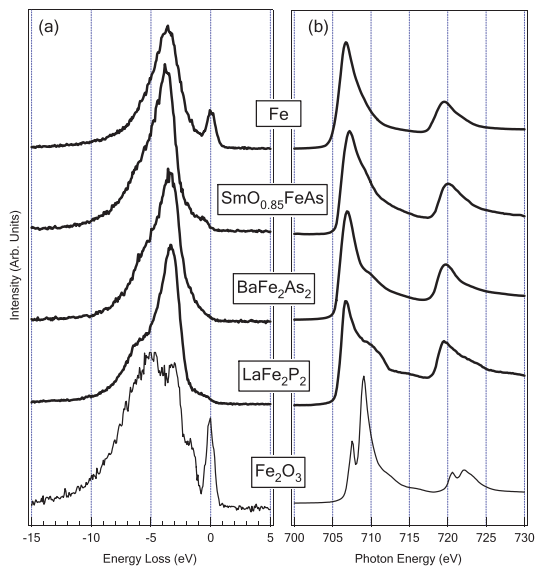}
\caption{\label{Fig:6}
(a) Comparison of RIXS data on the noted samples at 708eV excitation energy. All iron pnictides and iron metal show only fluorescent peaks same as the nonresonant spectrum (insets of Fig.1-4b), while $\alpha$-Fe$_2$O$_3$ displays multiple peaks as the signature of {\it dd}-excitations. (b) Comparison of XAS data on the same five samples, $\alpha$-Fe$_2$O$_3$ again displays very different lineshape due to the crystal field splitting.}
\end{figure}

For a direct comparison between all the samples, and for comparing with the theoretical calculations, Fig.6 shows the RIXS data collected with 708eV excitation energy as well as the XAS data on the five different samples. There are only minors difference in the symmetry of the lineshape and strength of the shoulders between the XAS data  of iron pnictides and iron metal; while the crystal field splitting leads to very different XAS spectrum of $\alpha$-Fe$_2$O$_3$ data (Fig.6b). The RIXS (Fig.6a) of iron pnictides and iron metal is dominated by the prominent fluorescent peak with no energy loss feature related to charge excitations; while for $\alpha$-Fe$_2$O$_3$, RIXS shows strong energy dependence and complex energy loss structure from electronic excitations and correlations. The similarity on the spectra between iron pnictides and iron metal, as well as the absence of charge excitation features in the RIXS data, suggests that iron pnictides are unlikely to be strongly correlated systems, which is further elaborated by theoretical calculations below.

\section{Calculations}
\label{Theory}

In order to understand the main features of the experimental spectra, namely that the spectra of the Fe pnictides greatly resemble that of Fe metal, we
proceed in three steps. First, to determine the importance of correlations, we present exact diagonalization calculations of a Hubbard model cluster 
which can be solved for either strong or weak Coulomb correlations. These calculations are used to give a qualitative estimate of the size of 
the Hubbard $U$ and Hund's $J$. Knowing the values of these parameters relative to the band width allows us to determine whether 
weakly-correlated methods are appropriate for describing the experimental spectra. Second, to better understand the relationship between multiplet,
spin-orbit, and crystal field effects on the spectra we present atomic multiplet calculations. And finally, having established the relatively minor
role of correlations in the Fe pnictides, we present DFT-based calculations of XAS and XES spectra for comparison with experiment. 

\subsection{Cluster Diagonalization}
\label{cluster}
In order to see explicitly how correlations affect the XAS profile, we perform a model many-body calculation based on the exact diagonalization (ED) technique, with a multi-orbital Hubbard model as the effective Hamiltonian. This approach has been successfully applied to understand the correlated physics in materials such as the high-T$_c$ cuprate parent compounds \cite{Vernay, Eskes, Kotani-Shin}.

The Fe pnictides have a tetrahedral FeAs$_4$ plaquette serving as the building block of the two dimensional Fe$_2$As$_2$ layer. We have therefore attempted to capture the essential physics revealed from XAS spectra with a FeAs$_4$ tetrahedral cluster including the five Fe 3$d$ levels and the As 4$p_{x,y,z}$ orbitals. The energy eigenstates that are necessary for calculating the XAS cross-sections \emph{via} Fermi's golden rule are then obtained by diagonalizing the multi-orbital Hamiltonian.

Our cluster calculations have been carried out in an assumed $d^6$ high-spin state for the Fe $3d$-levels, which is energetically preferred over the low-spin configuration due to Hund's coupling. While the experimentally measured magnetic moment in Fe pnictides is about $0.35\mu_B$, LDA predicts a larger magnetic moment \cite{Mazin}. This strength would decrease if the system is more delocalized.

The multi-orbital Hamiltonian entering the calculations can be written as $H=H_k+H_\epsilon+H_C+H_Q$. Here $H_k$ is the kinetic energy term:
%\begin{widetext}
\begin{eqnarray}\nonumber
H_k&=&\sum_{j,\gamma\gamma',\sigma} t_{pd,\gamma} (d_{\gamma\sigma}^\dagger p_{j\gamma'\sigma}+h.c.)
\\
&+&\sum_{jj',\gamma\gamma',\sigma} t_{pp,jj',\gamma\gamma'}(p_{j\gamma\sigma}^\dagger p_{j'\gamma'\sigma} + h.c.),
\end{eqnarray}
%\end{widetext}
where $d^\dagger_{\gamma\sigma}$ creates a particle with spin $\sigma$ in orbital $\gamma$ at the Fe site, and $p^\dagger_{j\gamma'\sigma}$ creates a particle with spin $\sigma$ in orbital $\gamma'$ at As site $j$. The relations among the multi-orbital hoppings are derived from the the Slater-Koster table \cite{Harrison}, with the strengths of these Slater-Koster matrix elements being $\vert V_{pd\sigma} \vert=$14.091(eV$\cdot \AA$) $\frac{\sqrt{r_p r_d^3}}{d^4}$, and $\vert V_{pd\pi}\vert=\frac{1}{\sqrt{3}} \vert V_{pd\sigma} \vert$. Here $d$ is the Fe-As bond length (in units of $\AA$), and the material specific values $r_p$ and $r_d$ are either given or can be calculated from Ref. \cite{Harrison}. In this work we use the
values: $d=2.39 \AA$, $r_p=13.2 \AA$, and $r_d=0.744 \AA$. We have further assumed that $\vert V_{pp\sigma}\vert=\frac{1}{2} \vert V_{pd\sigma} \vert$.

$H_\epsilon$ is the orbital site-energy term:
\begin{equation}
H_\epsilon= \sum_{\gamma\sigma}\epsilon_d(\gamma) n_{d,\gamma\sigma}+\sum_{j,\gamma\sigma} \epsilon_p n_{p,j\gamma\sigma},
\end{equation}
with $n_{d,\gamma\sigma}\equiv d^\dagger_{\gamma\sigma} d_{\gamma\sigma} $, and $n_{p,j\gamma\sigma}\equiv p^\dagger_{j\gamma\sigma} p_{j\gamma\sigma}$. The Fe $e_g$ and $t_{2g}$ orbital site energies are defined with respect to their center of gravity $\epsilon_d$ by $\epsilon_{d}(e_g)\equiv \epsilon_d-6Dq$, and $\epsilon_{d}(t_{2g})\equiv \epsilon_d+4Dq$. The arsenic $p$ orbital site energy $\epsilon_p$ is defined by $\Delta = \epsilon_d - \epsilon_p + n\overline{U}$ for the $d^n$ configuration, where $\Delta$ is the charge transfer gap energy. The subtraction of an average Coulomb repulsion term, $\overline{U}=A-\frac{14}{9}B+\frac{7}{9}C$, ensures a $d^n$ ground state; $A$, $B$, and $C$ are the Racah parameters.

The correlated physics is introduced directly from the Coulomb interaction term, including intra-orbital on-site Coulomb interactions, Hund's exchange coupling, and electron pair hopping processes, written as \cite{Dagotto}:
\begin{widetext}
\begin{eqnarray}
H_C=\frac{U}{2}\sum_{\gamma, \sigma\neq\sigma'} n_{d,\gamma\sigma}n_{d,\gamma\sigma'}+\frac{U'}{2}\sum_{\sigma,\sigma',\gamma\neq\gamma'}n_{d,\gamma\sigma}n_{d,\gamma'\sigma'}+\frac{J}{2}\sum_{\sigma, \sigma',\gamma\neq\gamma'} d_{\gamma\sigma}^\dagger d_{\gamma'\sigma'}^\dagger d_{\gamma\sigma'}d_{\gamma'\sigma}+\frac{J'}{2}\sum_{\sigma\neq\sigma',\gamma\neq\gamma'} d_{\gamma,\sigma}^\dagger d_{\gamma\sigma'}^\dagger d_{\gamma'\sigma'}d_{\gamma'\sigma}.
\end{eqnarray}
\end{widetext}
The above tight-binding parameters are related \emph{via} the Goodenough-Kanamori-Anderson relation: $U=U'+2J$, and $J=J'$. Written in terms of the Racah parameters, the on-site intra-orbital Coulomb repulsion $U$ is expressed as $U=A+4B+3C$. On the other hand, the Hund's coupling $J$, typically of order $\sim$ 1 eV in cuprates and other transition metal complexes, are solely determined by $B$ and $C$. Later we shall treat $A$ as an adjusting parameter, and hence the on-site Coulomb repulsions, seeing how it affects the XAS spectra.

For the XAS final states we include an additional core-hole potential term in the Hamiltonian:
\begin{equation}
H_Q=\sum_{\gamma,\sigma}U_Q n_{d,\gamma\sigma }n^c,
\end{equation}
where $n^c\equiv d^\dagger_c d_c$, and $d^\dagger_c$ is the creation operator for a hole in the core-hole orbital. The strength of $U_Q$ can be determined experimentally from the energy separation of the well-/poorly-screened resonances, and is of the same order of magnitude as $U$. Here we use $\vert U_Q \vert=U$ for simplicity. In short, the tight-binding parameters used in the calculations are (in units of eV): $V_{pd\sigma}=-1.10$, $V_{pd\pi}=0.63$, $V_{pp\sigma}=0.55$, and $V_{pp\pi}=-0.15$; $B=0.10$, and $C=0.40$ (resulting in a $J(e_g)=0.8$); $\epsilon_d\equiv 0.00$, $10Dq$=0.20, and $\Delta=1.50$.

\begin{figure}[h!]
\includegraphics[width=2.5in]{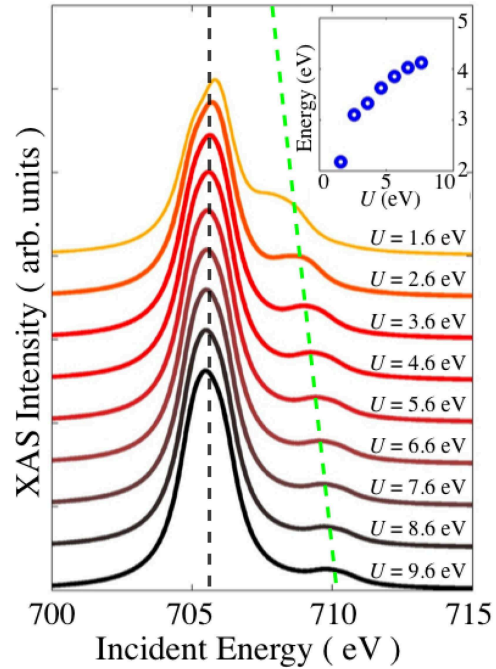}
 \caption{Fe-pnictide $L_3$-edge XAS spectra obtained from small cluster diagonalization for a fixed  $J(e_g)=0.8$ eV and various $U$. A strong Coulomb repulsion tends to suppress the XAS shoulder peak intensity. The dashed line sketches the energy separation of the dominant peak and its shoulder . The inset is a plot for the peak-shoulder energy separation versus the on-site repulsion $U$, from which a naive upper bound of $U\sim 2$ eV is drawn.}\label{fig:ED_U}
\end{figure}

\begin{figure}[h!]
\includegraphics[width=2.5in]{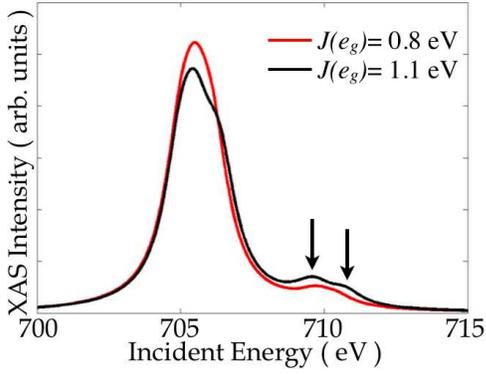}
\caption{Fe-pnictide $L_3$-edge XAS spectra obtained from small cluster diagonalization for a fixed  $U=8.6$ eV and two different values of $J(e_g)$. The XAS shoulder peak is split into a two-peak structure by a larger $J$, indicated by the black arrows.}\label{fig:ED_J}
\end{figure}

Fig. \ref{fig:ED_U} shows the Fe-pnictide $L_3$-edge XAS spectra from the cluster calculation, with varied Coulomb repulsion $U$. A stronger $U$ suppresses the XAS shoulder peak intensity. The shoulder peak is further split into a two-peak structure for larger Hund's coupling, as is shown in Fig. \ref{fig:ED_J}. By either increasing the covalency or reducing the correlation effects a more featureless XAS spectrum with a shoulder peak intensity comparable to experiments is obtained. A naive upper bound for the on-site repulsion is therefore drawn by looking at the XAS shoulder peak structure, as well as its energy separation from the dominant peak. According to the calculation, we estimate the Coulomb interactions to be $U\sim2$ eV and $J(e_g)=0.8$ eV, consistent with the empirical formula \cite{Marel}. This result suggests that it is more appropriate to treat Fe-pnictides as weakly-correlated systems.

A limitation of the cluster approach is that while it includes interaction between the 3d electrons explicitly, the states obtained from the cluster diagonalization do not include the atomic multiplet structures associated with the spin-orbit coupling of the Fe 2p core-hole. To test how this multiplet structure affects the resultant spectra, we have also computed XAS profiles including the atomic multiplets, at the expense of removing the pnictide atoms from the cluster.

\subsection{Multiplet Calculation}

\label{multiplet}
X-ray absorption spectra are calculated using Fermi's golden-rule, with a finite lifetime for the core-hole. Thus, the x-ray absorption intensity
may be written explicitly in terms of a sum over states $|{\psi_i}\rangle$ as
\begin{equation}\label{XAS_formula}
I(\omega)=\sum_i|\langle\psi_i|\hat{
%{\mathcal O}
d
}|\psi_0\rangle|^2
\frac{\Gamma/\pi}{(\omega+E_0-E_i)^2+\Gamma^2}\;.
\end{equation}  
%The RIXS process is second-order and Fermi's golden-rule yields  the well-known
%Kramers-Heisenberg formula;\cite{Kotani} after the absorption 
%a photon is emitted leaving the system in a final excited state $|\psi_f\rangle$. 

We considered the specific case of XAS experiments on BaFe$_2$As$_2$.  
During the optical transitions, the states  $|\psi_0\rangle$, $|\psi_i\rangle$
and $|\psi_f\rangle$ belong respectively to the configurations 
$2p^6 3d^6$, $2p^5 3d^7$ and $(2p^6 3d^6)^\star$. For each of these
configurations, we have to consider electron-electron interactions, spin-orbit and crystal-field 
on an equal footing while the radial wave-functions are determined by solving
the Dirac equation. This leads to a splitting of the shells into {\it multiplet levels}. 
To be specific, the crystal-field is expressed as an electronic potential of 
external point-charge ions interacting with the considered Fe-ion. 
From there we can evaluate the associated XAS spectra within the
dipole approximation.

\begin{figure}
\includegraphics[width=2.25in]{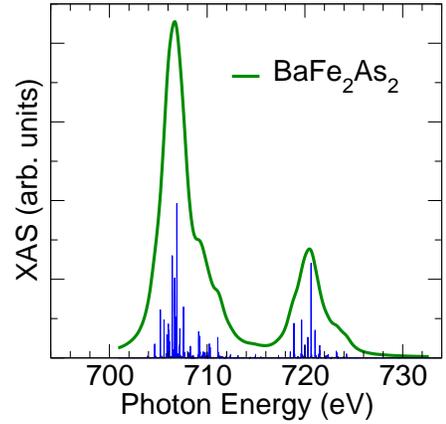}
 \caption{\label{Fig:XAS_multiplet}
X-ray absorption spectrum: Iron L-edge multiplet calculation for 
BaFe$_2$As$_2$.\\ }
\end{figure}

The obtained XAS spectrum of Fig.\ref{Fig:XAS_multiplet} shows good agreement with the experimental data and mostly exhibits the L$_2$-L$_3$ splitting coming from the 
spin-orbit splitting of the $2p$ core-levels. 
%In Fig.\ref{Fig:RIXS_multiplet}, in order to compare with the experimental RIXS data, we considered an in-plane polarization for the incoming photons 
%and used a constant broadening of 0.2eV in the energy-loss direction. 
%We can see that our result is qualitatively in good agreement with the experimental spectrum where we see a single asymmetric peak with a $\sim 10$eV %width, while in Fig.\ref{Fig:RIXS_multiplet} we resolve different modes within that energy range. 
However, the present calculation does not 
involve charge fluctuations, the explicit inclusion of these would give rise to satellite peaks in a RIXS spectrum 
and the inclusion of the ligands would give rise to the formation of bands where 
electron-electron scattering would occur. 
%these effects leading to corrections of the RIXS profile of Fig.\ref{Fig:RIXS_multiplet} and the merging of the different modes into a big asymmetric %peak. 
%\begin{figure}[h!]
%
%\begin{figure}[h!]
%\includegraphics[width=2.25in]{BaFe2As2_RIXS_plane.eps}
% \caption{\label{Fig:RIXS_multiplet}
%RIXS spectrum: multiplet calculation for BaFe$_2$As$_2$ for an incoming photon
% energy of 708eV and in-plane polarization.}
%\end{figure}

In the limit of a strongly correlated regime 
%($U\gg W$), 
the initial and final states cannot be described by a single-site approach, hence 
the XAS spectra would exhibit additional peaks that cannot be 
captured by the present multiplet picture. The experimental data do not exhibit such a signature: besides the spin-orbit splitting the XAS spectrum is rather featureless and the additional shoulders can be explained by the effects of the crystal-field on the atomic-multiplet. 
%The measured RIXS spectrum consists in a broad single peak while the calculated spectrum 
%exhibits different peaks. This discrepancy can be assigned to charge fluctuations and formation of bands, as discussed in the previous paragraph, which %are not taken into account in the present simulation. 
These facts combine to suggest that BaFe$_2$As$_2$ is rather not a strongly correlated material. 

\subsection{DFT-based FEFF Calculation}
\label{FEFF}
For weakly correlated materials XAS and XES spectra can reliably be computed using {\it ab initio} methods. Such calculations, however, have a considerable degree of complexity because in XAS/XES a core-hole is present in the final/initial state. The computer code {\tt FEFF}\cite{FEFF} is well-known for treating the core-hole potential with a high level of accuracy. Here we have used {\tt FEFF} to calculate the XAS near the $L_2$ and $L_3$ iron edges, the angular-momentum projected density of states (LDOS), and the XES spectra for the $L_3$ edge. 

Our calculations begin by overlapping relativistic Dirac-Fock atomic potentials via the Mattheiss prescription\cite{Mattheiss,Ank97}. This prescription fixes a ``Norman" radius about each atomic site which contains $Z_i$ electrons, where $Z_i$ is the atomic number of the atom at site $i$. The overlapped Mattheiss potentials are then used as the starting point of a self-consistent (SCF) potential calculation which uses the ground-state von Barth-Hedin exchange-correlation potential\cite{vonBarth} on all iterations. Given the SCF potential, the relativistic radial wave-functions and phase shifts associated with each atomic scattering site can be calculated. The single-electron
Green's function for the entire system may then be written using a basis of these radial wavefunctions and spherical harmonics with system (cluster) dependent coefficients that are calculated within multiple scattering theory.\cite{rehralbers} 
Finally, the XAS can be calculated from Eq.~(\ref{XAS_formula}),
using the Green's function to implicitly sum over states: 
\begin{equation}\label{Eq:XAS}
\mu\sim -{\rm Im}\langle{\psi_0}|\hat d^\dagger \hat G(E_c+\hbar\omega) \hat d |{\psi_0}\rangle \;,
\end{equation}
where $d$ is the single-electron dipole operator, $G$ is the photoelectron Green's function, 
and the state $|\psi_0\rangle$ is the core state of interest which, in this work, is either the ${\rm L}_2$ or ${\rm L}_3$ edge of iron. The energy of the absorbed photon is $\hbar \omega$, and the $\sim$ symbol means that we have neglected to write a number of constant prefactors as well as a broadened step function limiting the XAS spectrum to $\omega > |E_c|+\mu$, where $\mu$ is the Fermi level. The XES can be calculated from a similar formula, but is limited by the complementary step function to the absorption case. 
The presence of the core-hole on the absorbing atom, as well as the effect of the Hedin-Lundqvist\cite{HL} self-energy (in the ``plasmon pole"\cite{lundqvist} approximation), are also included in the {\tt FEFF} calculations.\cite{jose} The LDOS for each type of atom is calculated by integrating the spacially- and energy-dependent density about each atom within the Norman sphere. This allows for an unambiguous definition of the LDOS for each type of atom.% What is common in the FEFF calculations presented here is the domination of the near-edge XAS/XES spectra by the Fe d-projected DOS (dDOS).

\begin{figure}
\includegraphics[width=3.0in]{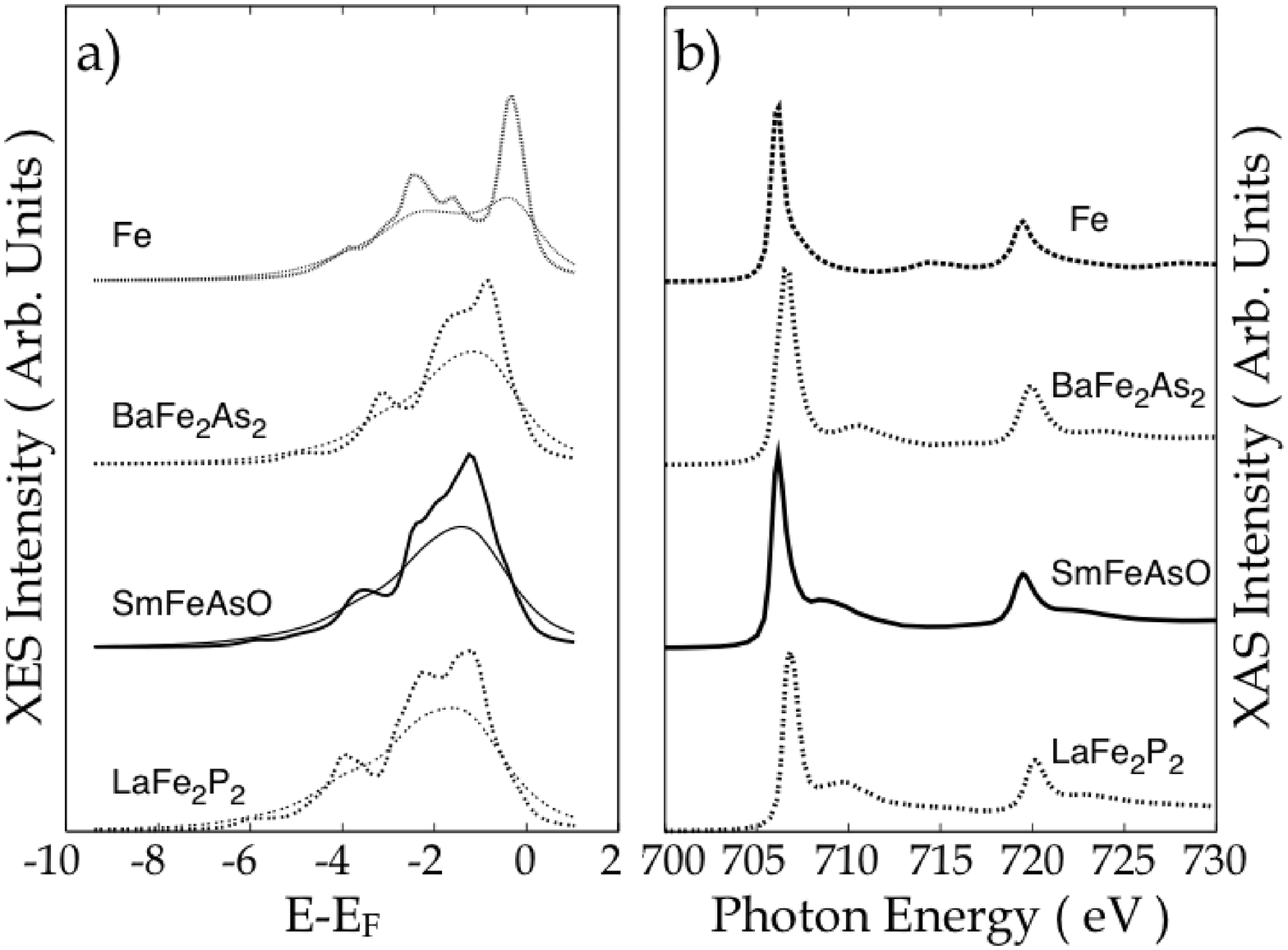} 
\caption{\label{Fig:FEFF}
The Fe $L_3$ edge XES signal (a) calculated using {\tt FEFF} for the four different metallic iron-containing materials considered in the experimental section. The Fe ${L}_{2,3}$ edge XAS (b) calculated using {\tt FEFF} for the four different iron-containing materials considered in the experimental section. 
}
\end{figure}

\begin{figure}
\includegraphics[width=2.5in]{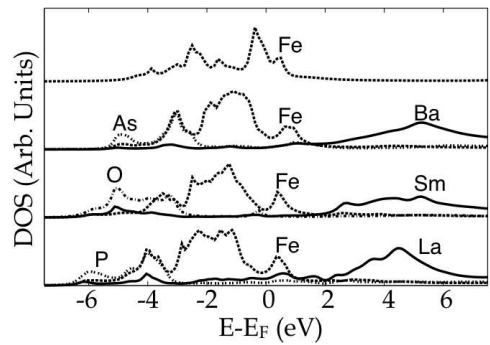} 
 \caption{The local angular-momentum projected densities of states for each of the metallic materials considered in the experimental section. Only the largest angular-momentum contributions from each type of atom are shown; the p-DOS is shown for As, O, and P; the d-DOS is shown for Ba, La, Sm, and Fe.
We note that the Fe DOS shown is that of the absorbing atom with the core hole.}\label{Fig:FEFF_DOS}
\end{figure}

In the panel b) of Fig.~\ref{Fig:FEFF} we show our {\tt FEFF} calculations of the iron ${L}_{2,3}$ edge XAS for the four metallic iron-containing materials considered in the experimental section. In the panel a) of Fig.~\ref{Fig:FEFF} we show our XES calculations at the iron $L_3$ edge. 
These XES calculations may be compared to the RIXS data for high fixed incoming energy and detected photon energies below $|E_{L_3}|+\mu$. Similarly to our interpretation of the XAS as reflecting the unoccupied dDOS, the near edge XES may be simply interpreted as a reflection of the 
occupied dDOS which is dominated by the Fe contribution near the Fermi level. 
%It is seen from the figure that the FEFF calculations agree well with experiment for four of the five materials, with the exception being the insulating iron oxide Fe$_2$O$_3$. 
%From this one may conclude that the quiddity of the quartet of materials other than Fe$_2$O$_3$ is their quality of metallicity. 
The similarity of the pnictide spectra to that of ordinary Fe metal underscores the importance of metalicity in these materials. 
%In 
%Fig.~(\ref{Fig:FEFF2}) we show the iron ${\rm L}_3$ edge XAS for the high-temperature phase of ${\rm Ba}{\rm Fe}_2{\rm As}_2$ as well as the d-projected 
%density of states (dDOS) of the absorbing iron atom. 
The calculated spectra in both the pnictides and iron metal mimic the DFT density of states
as expected in a weakly-correlated picture; the XAS (which involves an initial p-state) 
is determined in the near edge region largely by the unoccupied dDOS. 
The dDOS, in turn, is dominated near the Fermi level by the contribution from iron for all the materials considered, as shown in Fig.~\ref{Fig:FEFF_DOS}. 
Other features of both the XAS and the XES can be matched to the peaks in the dDOS shown in Fig.~\ref{Fig:FEFF_DOS}. For completeness we
also show the pDOS of As, O, and P, in Fig.~\ref{Fig:FEFF_DOS}, but we note that it is not directly related to the XAS or XES spectra.

%brian's figure
\begin{figure}
\includegraphics[width=3.0in]{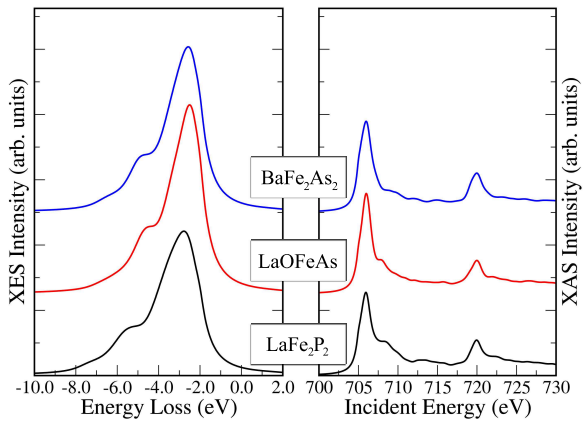}
\caption{\label{Fig:Bri}
The iron ${\rm L}_{2,3}$ edge XAS and XES, calculated using {\tt WIEN2k}, for three different iron containing materials.}
\end{figure}

The fact that in {\tt FEFF} the features of the XAS and the XES spectra can be matched to the peaks in the Fe 3d densities of states suggests that the core-hole is relatively unimportant and its potential weak. We can test this conclusion by computing the same spectra with a method that disregards the potential of the 2p core-hole, but retains the correct band-structure dipole matrix elements. For this we used the plane-wave based DFT computer program {\tt WIEN2k}\cite{wien}. The resulting XES and XAS spectra for three types of iron based materials are shown in Fig.~\ref{Fig:Bri}. The $L_2$-$L_3$ splitting and relative intensity cannot be determined with this method and are therefore introduced by hand. The computed XAS and XES spectra of the pnictides agree well with both our {\tt FEFF} calculations and the experiments. We observed that in both {\tt FEFF} and {\tt WIEN2k} calculations, the effect of oxygen doping and also the effect of replacement of one rare earth with another is rather small since the XAS is determined largely by the local environment of the absorber which is dominated by iron. We also note that the theoretical spectra presented here are polarization averaged; we have investigated polarization dependence in our {\tt FEFF} calculations but find no significant differences between polarized and averaged spectra in the pnictides.  

\section{Summary}
\label{summary}

In the young field of iron-pnictide superconductors there are currently several open issues. For instance, it is not well understood why the observed ordered magnetic moment is so small. Also the role of local Coulomb interactions is not well characterized, as is the degree of As hybridization with Fe orbitals near the Fermi level. The tendency towards the formation of large iron moments due to local Hund's rule exchange and a possible emerging role for orbital degrees of freedom are other disputed issues~\cite{Johannes,Kruger09,Hozoi09,Singh}. These disputes stand in the way of a consensus on the minimal model needed to describe the physics and ultimately the pairing mechanism in Fe-pnictide superconductors. In this context we have investigated five iron containing materials including 122 and 1111 Fe-pnictides with a combination of XAS and RIXS techniques. The first general observation is that the experimental data for the Fe pnictides is qualitatively similar to other metallic Fe materials and significantly different from large gap Fe-based insulators such as hematite $\alpha$-Fe$_2$O$_3$. 

The three main theoretical approaches that we have used to analyze the data incorporate electronic correlation effects due to electron-electron interactions to different degrees. If the pnictides were very localized one would expect that the essential features of XAS and RIXS spectra can readily be captured in a small FeAs cluster. Our exact diagonalization computation of the absorption spectra for such a cluster in the localized, strong coupling limit (large Hubbard $U$) clearly shows the appearance of a high energy peak well separated from the main absorption line at the L$_3$ edge. In the experimental data this peak is absent -- or rather appears as a shoulder of the main XAS line. From a comparison of the energy position of this shoulder in the data and the cluster simulation we extract an upper limit of the Hubbard $U$ of 2 eV --substantially smaller than the Fe $3d$ bandwidth. Hund's rule $J$ is about 0.8 eV. Coulomb correlations are thus much weaker then in the cuprates. This result is confirmed by our multiplet calculation, which is a local approach that has the advantage of including the spin-orbit and Coulomb interactions related to the core-hole. The calculated $L_3$-$L_2$ edge energy splittings and intensity ratios agree with the experimental data. 

The inference that the Hubbard $U$ is small and the $3d$ electrons weakly correlated, suggests a comparison of the spectroscopic data with the results of single-particle, {\it ab initio} calculations. Such approaches are complicated by the fact that a core-hole is present in the final state of XAS and the intermediate state of RIXS. The Coulomb interaction of the core-hole with the valence electrons can in principle be strong and such an electronic correlation effect has a profound influence on XAS and RIXS spectra. The density-functional based {\tt FEFF} code treats the effects of the core-hole potential with a high level of accuracy.  The spectra computed with {\tt FEFF} confirm the presence of the XAS shoulder and agree with the experimental XAS and RIXS data very well. We have used this agreement to further filter out the core-hole induced correlations. When calculating the spectra with plane-wave based {\tt WIEN2k} code --which includes the proper dipole transition matrix elements, but lacks the final state XAS core-hole potential-- we observe that the simulated spectra basically do not change. We conclude that in the Fe pnictides not only the Hubbard $U$ but also the core-hole potential is therefore heavily screened. The electronic correlations that the core-hole induces are thus weak and consequently the spectra can safely be interpreted in terms of single-particle densities of states and the appropriate dipole transition matrix elements. The present spectroscopic data and its theoretical description thus emphasize the role of strong covalency and Fe metalicity in the Fe pnictides.

\acknowledgments
The authors would like to acknowledge important discussions with J. Zaanen, I. Mazin, D. Reznik, J. J. Rehr, A. Baron, S. Johnston, Y.-D. Chuang, W. A. Harrison, S. Kumar and M. Golden.
This work is supported by the Office of Science of the U.S. Department of Energy (DOE) under Contract No. DE-AC02-76SF00515, and DE-FG02-08ER4650 (CMSN). This research used resources of the National Energy Research Scientific Computing Center, which is supported by DOE under Contract No. DE-AC02-05CH11231.
This work is supported by the ``Stichting voor Fundamenteel Onderzoek der Materie (FOM)".  The Advanced Light Source (ALS) is supported by the Director, Office of Science, Office of Basic Energy Sciences, of the U.S. Department of Energy under Contract No. DE-AC02-05CH11231. P.~O.~would like to 
acknowledge the support from CONTACyT, Mexico.

%\bibliography{mybib}{}

\begin{thebibliography}{55}
\expandafter\ifx\csname natexlab\endcsname\relax\def\natexlab#1{#1}\fi
\expandafter\ifx\csname bibnamefont\endcsname\relax
  \def\bibnamefont#1{#1}\fi
\expandafter\ifx\csname bibfnamefont\endcsname\relax
  \def\bibfnamefont#1{#1}\fi
\expandafter\ifx\csname citenamefont\endcsname\relax
  \def\citenamefont#1{#1}\fi
\expandafter\ifx\csname url\endcsname\relax
  \def\url#1{\texttt{#1}}\fi
\expandafter\ifx\csname urlprefix\endcsname\relax\def\urlprefix{URL }\fi
\providecommand{\bibinfo}[2]{#2}
\providecommand{\eprint}[2][]{\url{#2}}

\bibitem[{\citenamefont{Boeri et~al.}(2008)\citenamefont{Boeri, Golgov, and
  Golubov}}]{eph1}
\bibinfo{author}{\bibfnamefont{L.}~\bibnamefont{Boeri}},
  \bibinfo{author}{\bibfnamefont{O.~V.} \bibnamefont{Golgov}},
  \bibnamefont{and} \bibinfo{author}{\bibfnamefont{A.~A.}
  \bibnamefont{Golubov}}, \bibinfo{journal}{Phys. Rev. Lett.}
  \textbf{\bibinfo{volume}{101}}, \bibinfo{pages}{026403}
  (\bibinfo{year}{2008}).

\bibitem[{\citenamefont{Reznik et~al.}(2008)\citenamefont{Reznik, Lokshin,
  Mitchell, Parshall, Dmowski, Lamago, Heid, K.-P-Bohnen, Sefat, McGuire
  et~al.}}]{eph2}
\bibinfo{author}{\bibfnamefont{D.}~\bibnamefont{Reznik}},
  \bibinfo{author}{\bibfnamefont{K.}~\bibnamefont{Lokshin}},
  \bibinfo{author}{\bibfnamefont{D.~C.} \bibnamefont{Mitchell}},
  \bibinfo{author}{\bibfnamefont{D.}~\bibnamefont{Parshall}},
  \bibinfo{author}{\bibfnamefont{W.}~\bibnamefont{Dmowski}},
  \bibinfo{author}{\bibfnamefont{D.}~\bibnamefont{Lamago}},
  \bibinfo{author}{\bibfnamefont{R.}~\bibnamefont{Heid}},
  \bibinfo{author}{\bibnamefont{K.-P-Bohnen}},
  \bibinfo{author}{\bibfnamefont{A.~S.} \bibnamefont{Sefat}},
  \bibinfo{author}{\bibfnamefont{M.~A.} \bibnamefont{McGuire}},
  \bibnamefont{et~al.}, \bibinfo{journal}{arXiv:0810.4941}
  (\bibinfo{year}{2008}).

\bibitem[{\citenamefont{Zhang et~al.}(2008)\citenamefont{Zhang, Oyanagi, Sun,
  Kamihara, and Hosono}}]{eph3}
\bibinfo{author}{\bibfnamefont{C.~J.} \bibnamefont{Zhang}},
  \bibinfo{author}{\bibfnamefont{H.}~\bibnamefont{Oyanagi}},
  \bibinfo{author}{\bibfnamefont{Z.~H.} \bibnamefont{Sun}},
  \bibinfo{author}{\bibfnamefont{Y.}~\bibnamefont{Kamihara}}, \bibnamefont{and}
  \bibinfo{author}{\bibfnamefont{H.}~\bibnamefont{Hosono}},
  \bibinfo{journal}{Phys. Rev. B} \textbf{\bibinfo{volume}{78}},
  \bibinfo{pages}{214513} (\bibinfo{year}{2008}).

\bibitem[{\citenamefont{Kivelson and Yao}(2008)}]{Kivelson}
\bibinfo{author}{\bibfnamefont{S.~A.} \bibnamefont{Kivelson}} \bibnamefont{and}
  \bibinfo{author}{\bibfnamefont{H.}~\bibnamefont{Yao}},
  \bibinfo{journal}{Nature Materials} \textbf{\bibinfo{volume}{7}},
  \bibinfo{pages}{927} (\bibinfo{year}{2008}).

\bibitem[{\citenamefont{Zhai et~al.}(2008)\citenamefont{Zhai, Wang, and
  Lee}}]{DHLee}
\bibinfo{author}{\bibfnamefont{H.}~\bibnamefont{Zhai}},
  \bibinfo{author}{\bibfnamefont{F.}~\bibnamefont{Wang}}, \bibnamefont{and}
  \bibinfo{author}{\bibfnamefont{D.-H.} \bibnamefont{Lee}},
  \bibinfo{journal}{arXiv:0810.2320}  (\bibinfo{year}{2008}).

\bibitem[{\citenamefont{Kuroki et~al.}(2008)\citenamefont{Kuroki, Onari, Arita,
  Usui, Tanaka, Kontani, and Aoki}}]{Graser1}
\bibinfo{author}{\bibfnamefont{K.}~\bibnamefont{Kuroki}},
  \bibinfo{author}{\bibfnamefont{S.}~\bibnamefont{Onari}},
  \bibinfo{author}{\bibfnamefont{R.}~\bibnamefont{Arita}},
  \bibinfo{author}{\bibfnamefont{H.}~\bibnamefont{Usui}},
  \bibinfo{author}{\bibfnamefont{Y.}~\bibnamefont{Tanaka}},
  \bibinfo{author}{\bibfnamefont{H.}~\bibnamefont{Kontani}}, \bibnamefont{and}
  \bibinfo{author}{\bibfnamefont{H.}~\bibnamefont{Aoki}},
  \bibinfo{journal}{Phys. Rev. Lett.} \textbf{\bibinfo{volume}{101}},
  \bibinfo{pages}{087004} (\bibinfo{year}{2008}).

\bibitem[{\citenamefont{Graser et~al.}(2008)\citenamefont{Graser, Maier,
  Hirschfeld, and Scalapino}}]{Graser2}
\bibinfo{author}{\bibfnamefont{S.}~\bibnamefont{Graser}},
  \bibinfo{author}{\bibfnamefont{T.~A.} \bibnamefont{Maier}},
  \bibinfo{author}{\bibfnamefont{P.~J.} \bibnamefont{Hirschfeld}},
  \bibnamefont{and} \bibinfo{author}{\bibfnamefont{D.~J.}
  \bibnamefont{Scalapino}}, \bibinfo{journal}{arXiv:0812.0343}
  (\bibinfo{year}{2008}).

\bibitem[{\citenamefont{Mazin et~al.}(2008)\citenamefont{Mazin, Johannes,
  Boeri, Koepernik, and Singh}}]{Mazin}
\bibinfo{author}{\bibfnamefont{I.}~\bibnamefont{Mazin}},
  \bibinfo{author}{\bibfnamefont{M.~D.} \bibnamefont{Johannes}},
  \bibinfo{author}{\bibfnamefont{L.}~\bibnamefont{Boeri}},
  \bibinfo{author}{\bibfnamefont{K.}~\bibnamefont{Koepernik}},
  \bibnamefont{and} \bibinfo{author}{\bibfnamefont{D.~J.} \bibnamefont{Singh}},
  \bibinfo{journal}{Phys. Rev. B} \textbf{\bibinfo{volume}{78}},
  \bibinfo{pages}{085104} (\bibinfo{year}{2008}).

\bibitem[{\citenamefont{Giovannetti et~al.}(2008)\citenamefont{Giovannetti,
  Kumar, and van~den Brink}}]{Giovannetti08}
\bibinfo{author}{\bibfnamefont{G.}~\bibnamefont{Giovannetti}},
  \bibinfo{author}{\bibfnamefont{S.}~\bibnamefont{Kumar}}, \bibnamefont{and}
  \bibinfo{author}{\bibfnamefont{J.}~\bibnamefont{van~den Brink}},
  \bibinfo{journal}{Phys. B} \textbf{\bibinfo{volume}{403}},
  \bibinfo{pages}{3653} (\bibinfo{year}{2008}).

\bibitem[{\citenamefont{Haule et~al.}(2008)\citenamefont{Haule, Shim, and
  Kotliar}}]{Haule}
\bibinfo{author}{\bibfnamefont{K.}~\bibnamefont{Haule}},
  \bibinfo{author}{\bibfnamefont{J.~H.} \bibnamefont{Shim}}, \bibnamefont{and}
  \bibinfo{author}{\bibfnamefont{G.}~\bibnamefont{Kotliar}},
  \bibinfo{journal}{Phys. Rev. Lett.} \textbf{\bibinfo{volume}{100}},
  \bibinfo{pages}{226402} (\bibinfo{year}{2008}).

\bibitem[{\citenamefont{Craco et~al.}(2008)\citenamefont{Craco, Laad, Leoni,
  and Rosner}}]{Craco}
\bibinfo{author}{\bibfnamefont{L.}~\bibnamefont{Craco}},
  \bibinfo{author}{\bibfnamefont{M.~S.} \bibnamefont{Laad}},
  \bibinfo{author}{\bibfnamefont{S.}~\bibnamefont{Leoni}}, \bibnamefont{and}
  \bibinfo{author}{\bibfnamefont{H.}~\bibnamefont{Rosner}},
  \bibinfo{journal}{Phys. Rev. B} \textbf{\bibinfo{volume}{78}},
  \bibinfo{pages}{134511} (\bibinfo{year}{2008}).

\bibitem[{\citenamefont{Vildosola et~al.}(2008)\citenamefont{Vildosola,
  Pourovskii, Arita, Biermann, and Georges}}]{Georges}
\bibinfo{author}{\bibfnamefont{V.}~\bibnamefont{Vildosola}},
  \bibinfo{author}{\bibfnamefont{L.}~\bibnamefont{Pourovskii}},
  \bibinfo{author}{\bibfnamefont{R.}~\bibnamefont{Arita}},
  \bibinfo{author}{\bibfnamefont{S.}~\bibnamefont{Biermann}}, \bibnamefont{and}
  \bibinfo{author}{\bibfnamefont{A.}~\bibnamefont{Georges}},
  \bibinfo{journal}{Phys. Rev. B} \textbf{\bibinfo{volume}{78}},
  \bibinfo{pages}{064518} (\bibinfo{year}{2008}).

\bibitem[{\citenamefont{Anisimov
  et~al.}(2008{\natexlab{a}})\citenamefont{Anisimov, Korotin, Streltsov,
  Kozhevnikov, Kunes, Shorikov, and Korotin}}]{Anisimov1}
\bibinfo{author}{\bibfnamefont{V.~I.} \bibnamefont{Anisimov}},
  \bibinfo{author}{\bibfnamefont{D.~M.} \bibnamefont{Korotin}},
  \bibinfo{author}{\bibfnamefont{S.~V.} \bibnamefont{Streltsov}},
  \bibinfo{author}{\bibfnamefont{A.~V.} \bibnamefont{Kozhevnikov}},
  \bibinfo{author}{\bibfnamefont{J.}~\bibnamefont{Kunes}},
  \bibinfo{author}{\bibfnamefont{A.~O.} \bibnamefont{Shorikov}},
  \bibnamefont{and} \bibinfo{author}{\bibfnamefont{M.~A.}
  \bibnamefont{Korotin}}, \bibinfo{journal}{arXiv:0807.0547}
  (\bibinfo{year}{2008}{\natexlab{a}}).

\bibitem[{\citenamefont{Shorikov et~al.}(2008)\citenamefont{Shorikov, Korotin,
  Streltsov, Korotin, Anisimov, and Skornyakov}}]{Anisimov2}
\bibinfo{author}{\bibfnamefont{A.}~\bibnamefont{Shorikov}},
  \bibinfo{author}{\bibfnamefont{M.}~\bibnamefont{Korotin}},
  \bibinfo{author}{\bibfnamefont{S.}~\bibnamefont{Streltsov}},
  \bibinfo{author}{\bibfnamefont{D.}~\bibnamefont{Korotin}},
  \bibinfo{author}{\bibfnamefont{V.}~\bibnamefont{Anisimov}}, \bibnamefont{and}
  \bibinfo{author}{\bibfnamefont{S.}~\bibnamefont{Skornyakov}},
  \bibinfo{journal}{arXiv:0804.3283}  (\bibinfo{year}{2008}).

\bibitem[{\citenamefont{Anisimov
  et~al.}(2008{\natexlab{b}})\citenamefont{Anisimov, Korotin, Korotin,
  Kozhevnikov, Kunes, Shorikov, Skornyakov, and Streltsov}}]{Anisimov3}
\bibinfo{author}{\bibfnamefont{V.~I.} \bibnamefont{Anisimov}},
  \bibinfo{author}{\bibfnamefont{D.~M.} \bibnamefont{Korotin}},
  \bibinfo{author}{\bibfnamefont{M.~A.} \bibnamefont{Korotin}},
  \bibinfo{author}{\bibfnamefont{A.~V.} \bibnamefont{Kozhevnikov}},
  \bibinfo{author}{\bibfnamefont{J.}~\bibnamefont{Kunes}},
  \bibinfo{author}{\bibfnamefont{A.~O.} \bibnamefont{Shorikov}},
  \bibinfo{author}{\bibfnamefont{S.}~\bibnamefont{Skornyakov}},
  \bibnamefont{and} \bibinfo{author}{\bibfnamefont{S.~V.}
  \bibnamefont{Streltsov}}, \bibinfo{journal}{arXiv:0810.2629}
  (\bibinfo{year}{2008}{\natexlab{b}}).

\bibitem[{\citenamefont{Lu et~al.}(2008)\citenamefont{Lu, Yi, Mo, Erickson,
  Analytis, Chu, Singh, Hussain, Geballe, Fisher et~al.}}]{Lu}
\bibinfo{author}{\bibfnamefont{D.~H.} \bibnamefont{Lu}},
  \bibinfo{author}{\bibfnamefont{M.}~\bibnamefont{Yi}},
  \bibinfo{author}{\bibfnamefont{S.-K.} \bibnamefont{Mo}},
  \bibinfo{author}{\bibfnamefont{A.~S.} \bibnamefont{Erickson}},
  \bibinfo{author}{\bibfnamefont{J.}~\bibnamefont{Analytis}},
  \bibinfo{author}{\bibfnamefont{J.-H.} \bibnamefont{Chu}},
  \bibinfo{author}{\bibfnamefont{D.~J.} \bibnamefont{Singh}},
  \bibinfo{author}{\bibfnamefont{Z.}~\bibnamefont{Hussain}},
  \bibinfo{author}{\bibfnamefont{T.~H.} \bibnamefont{Geballe}},
  \bibinfo{author}{\bibfnamefont{I.~R.} \bibnamefont{Fisher}},
  \bibnamefont{et~al.}, \bibinfo{journal}{Nature}
  \textbf{\bibinfo{volume}{455}}, \bibinfo{pages}{81} (\bibinfo{year}{2008}).

\bibitem[{\citenamefont{de~Groot and Kotani}(2008)}]{Degroot}
\bibinfo{author}{\bibfnamefont{F.}~\bibnamefont{de~Groot}} \bibnamefont{and}
  \bibinfo{author}{\bibfnamefont{A.}~\bibnamefont{Kotani}},
  \emph{\bibinfo{title}{Core Level Spectroscopy of Solids}}
  (\bibinfo{publisher}{CRC Press}, \bibinfo{year}{2008}).

\bibitem[{\citenamefont{Kroll et~al.}(2008)\citenamefont{Kroll, Bonhommeau,
  Kachel, Duerr, Werner, Behr, Koitzsch, Huebel, Leger, Schoenfelder
  et~al.}}]{Kroll}
\bibinfo{author}{\bibfnamefont{T.}~\bibnamefont{Kroll}},
  \bibinfo{author}{\bibfnamefont{S.}~\bibnamefont{Bonhommeau}},
  \bibinfo{author}{\bibfnamefont{T.}~\bibnamefont{Kachel}},
  \bibinfo{author}{\bibfnamefont{H.~A.} \bibnamefont{Duerr}},
  \bibinfo{author}{\bibfnamefont{J.}~\bibnamefont{Werner}},
  \bibinfo{author}{\bibfnamefont{G.}~\bibnamefont{Behr}},
  \bibinfo{author}{\bibfnamefont{A.}~\bibnamefont{Koitzsch}},
  \bibinfo{author}{\bibfnamefont{R.}~\bibnamefont{Huebel}},
  \bibinfo{author}{\bibfnamefont{S.}~\bibnamefont{Leger}},
  \bibinfo{author}{\bibfnamefont{R.}~\bibnamefont{Schoenfelder}},
  \bibnamefont{et~al.}, \bibinfo{journal}{arXiv:0806.2625}
  (\bibinfo{year}{2008}).

\bibitem[{\citenamefont{Kurmaev et~al.}(2008)\citenamefont{Kurmaev, Wilks,
  Moewes, Skorikov, Izyumov, Finkelstein, Li, and Chen}}]{Moewes1}
\bibinfo{author}{\bibfnamefont{E.~Z.} \bibnamefont{Kurmaev}},
  \bibinfo{author}{\bibfnamefont{R.~G.} \bibnamefont{Wilks}},
  \bibinfo{author}{\bibfnamefont{A.}~\bibnamefont{Moewes}},
  \bibinfo{author}{\bibfnamefont{N.~A.} \bibnamefont{Skorikov}},
  \bibinfo{author}{\bibfnamefont{Y.~A.} \bibnamefont{Izyumov}},
  \bibinfo{author}{\bibfnamefont{L.~D.} \bibnamefont{Finkelstein}},
  \bibinfo{author}{\bibfnamefont{R.~H.} \bibnamefont{Li}}, \bibnamefont{and}
  \bibinfo{author}{\bibfnamefont{X.~H.} \bibnamefont{Chen}},
  \bibinfo{journal}{Phys. Rev. B} \textbf{\bibinfo{volume}{78}},
  \bibinfo{pages}{220503R} (\bibinfo{year}{2008}).

\bibitem[{\citenamefont{Kurmaev et~al.}(2009)\citenamefont{Kurmaev, McLeod,
  Buling, Skorikov, Moewes, Neumann, Korotin, Izyumov, Ni, and
  Canfield}}]{Moewes2}
\bibinfo{author}{\bibfnamefont{E.}~\bibnamefont{Kurmaev}},
  \bibinfo{author}{\bibfnamefont{J.}~\bibnamefont{McLeod}},
  \bibinfo{author}{\bibfnamefont{A.}~\bibnamefont{Buling}},
  \bibinfo{author}{\bibfnamefont{N.}~\bibnamefont{Skorikov}},
  \bibinfo{author}{\bibfnamefont{A.}~\bibnamefont{Moewes}},
  \bibinfo{author}{\bibfnamefont{M.}~\bibnamefont{Neumann}},
  \bibinfo{author}{\bibfnamefont{M.}~\bibnamefont{Korotin}},
  \bibinfo{author}{\bibfnamefont{Y.}~\bibnamefont{Izyumov}},
  \bibinfo{author}{\bibfnamefont{N.}~\bibnamefont{Ni}}, \bibnamefont{and}
  \bibinfo{author}{\bibfnamefont{P.}~\bibnamefont{Canfield}},
  \bibinfo{journal}{arXiv:09021141}  (\bibinfo{year}{2009}).

\bibitem[{\citenamefont{Ankudinov et~al.}(1998)\citenamefont{Ankudinov, Ravel,
  Rehr, and Conradson}}]{FEFF}
\bibinfo{author}{\bibfnamefont{A.~L.} \bibnamefont{Ankudinov}},
  \bibinfo{author}{\bibfnamefont{B.}~\bibnamefont{Ravel}},
  \bibinfo{author}{\bibfnamefont{J.~J.} \bibnamefont{Rehr}}, \bibnamefont{and}
  \bibinfo{author}{\bibfnamefont{S.~D.} \bibnamefont{Conradson}},
  \bibinfo{journal}{Phys. Rev. B} \textbf{\bibinfo{volume}{58}},
  \bibinfo{pages}{7565} (\bibinfo{year}{1998}).

\bibitem[{\citenamefont{Ren et~al.}(2008{\natexlab{a}})\citenamefont{Ren, Lu,
  Yang, Yi, Shen, Li, Che, Dong, Sun, Zhou et~al.}}]{ZhiAn1}
\bibinfo{author}{\bibfnamefont{Z.~A.} \bibnamefont{Ren}},
  \bibinfo{author}{\bibfnamefont{W.}~\bibnamefont{Lu}},
  \bibinfo{author}{\bibfnamefont{J.}~\bibnamefont{Yang}},
  \bibinfo{author}{\bibfnamefont{W.}~\bibnamefont{Yi}},
  \bibinfo{author}{\bibfnamefont{X.~L.} \bibnamefont{Shen}},
  \bibinfo{author}{\bibfnamefont{Z.~C.} \bibnamefont{Li}},
  \bibinfo{author}{\bibfnamefont{G.~C.} \bibnamefont{Che}},
  \bibinfo{author}{\bibfnamefont{X.~L.} \bibnamefont{Dong}},
  \bibinfo{author}{\bibfnamefont{L.~L.} \bibnamefont{Sun}},
  \bibinfo{author}{\bibfnamefont{F.}~\bibnamefont{Zhou}}, \bibnamefont{et~al.},
  \bibinfo{journal}{Chin. Phys. Lett.} \textbf{\bibinfo{volume}{25}},
  \bibinfo{pages}{2215} (\bibinfo{year}{2008}{\natexlab{a}}).

\bibitem[{\citenamefont{Ren et~al.}(2008{\natexlab{b}})\citenamefont{Ren, Yang,
  Lu, Yi, Shen, Li, Che, Dong, Sun, Zhou et~al.}}]{ZhiAn2}
\bibinfo{author}{\bibfnamefont{Z.~A.} \bibnamefont{Ren}},
  \bibinfo{author}{\bibfnamefont{J.}~\bibnamefont{Yang}},
  \bibinfo{author}{\bibfnamefont{W.}~\bibnamefont{Lu}},
  \bibinfo{author}{\bibfnamefont{W.}~\bibnamefont{Yi}},
  \bibinfo{author}{\bibfnamefont{X.~L.} \bibnamefont{Shen}},
  \bibinfo{author}{\bibfnamefont{Z.~C.} \bibnamefont{Li}},
  \bibinfo{author}{\bibfnamefont{G.~C.} \bibnamefont{Che}},
  \bibinfo{author}{\bibfnamefont{X.~L.} \bibnamefont{Dong}},
  \bibinfo{author}{\bibfnamefont{L.~L.} \bibnamefont{Sun}},
  \bibinfo{author}{\bibfnamefont{F.}~\bibnamefont{Zhou}}, \bibnamefont{et~al.},
  \bibinfo{journal}{EuroPhys. Lett.} \textbf{\bibinfo{volume}{82}},
  \bibinfo{pages}{57002} (\bibinfo{year}{2008}{\natexlab{b}}).

\bibitem[{\citenamefont{Ni et~al.}(2008)\citenamefont{Ni, Budko, Kreyssig,
  Nandi, Rustan, Goldman, Gupta, Corbett, Kracher, and
  Canfield}}]{Fishercrystals1}
\bibinfo{author}{\bibfnamefont{N.}~\bibnamefont{Ni}},
  \bibinfo{author}{\bibfnamefont{S.~L.} \bibnamefont{Budko}},
  \bibinfo{author}{\bibfnamefont{A.}~\bibnamefont{Kreyssig}},
  \bibinfo{author}{\bibfnamefont{S.}~\bibnamefont{Nandi}},
  \bibinfo{author}{\bibfnamefont{G.~E.} \bibnamefont{Rustan}},
  \bibinfo{author}{\bibfnamefont{A.~I.} \bibnamefont{Goldman}},
  \bibinfo{author}{\bibfnamefont{S.}~\bibnamefont{Gupta}},
  \bibinfo{author}{\bibfnamefont{J.~D.} \bibnamefont{Corbett}},
  \bibinfo{author}{\bibfnamefont{A.}~\bibnamefont{Kracher}}, \bibnamefont{and}
  \bibinfo{author}{\bibfnamefont{P.~C.} \bibnamefont{Canfield}},
  \bibinfo{journal}{Phys. Rev. B} \textbf{\bibinfo{volume}{78}},
  \bibinfo{pages}{014507} (\bibinfo{year}{2008}).

\bibitem[{\citenamefont{Chen et~al.}(2008)\citenamefont{Chen, Li, Dong, Li, Hu,
  Zhang, Song, Zheng, Wang, and Luo}}]{Fishercrystals2}
\bibinfo{author}{\bibfnamefont{G.~F.} \bibnamefont{Chen}},
  \bibinfo{author}{\bibfnamefont{Z.}~\bibnamefont{Li}},
  \bibinfo{author}{\bibfnamefont{J.}~\bibnamefont{Dong}},
  \bibinfo{author}{\bibfnamefont{G.}~\bibnamefont{Li}},
  \bibinfo{author}{\bibfnamefont{W.~Z.} \bibnamefont{Hu}},
  \bibinfo{author}{\bibfnamefont{X.~D.} \bibnamefont{Zhang}},
  \bibinfo{author}{\bibfnamefont{X.~H.} \bibnamefont{Song}},
  \bibinfo{author}{\bibfnamefont{P.}~\bibnamefont{Zheng}},
  \bibinfo{author}{\bibfnamefont{N.~L.} \bibnamefont{Wang}}, \bibnamefont{and}
  \bibinfo{author}{\bibfnamefont{J.~L.} \bibnamefont{Luo}},
  \bibinfo{journal}{arXiv:0806.2648}  (\bibinfo{year}{2008}).

\bibitem[{\citenamefont{Analytis et~al.}(2009)\citenamefont{Analytis, McDonald,
  Chu, Riggs, Bangura, Kucharczyk, Johannes, and Fisher}}]{fish}
\bibinfo{author}{\bibfnamefont{J.~G.} \bibnamefont{Analytis}},
  \bibinfo{author}{\bibfnamefont{R.~D.} \bibnamefont{McDonald}},
  \bibinfo{author}{\bibfnamefont{J.-H.} \bibnamefont{Chu}},
  \bibinfo{author}{\bibfnamefont{S.~C.} \bibnamefont{Riggs}},
  \bibinfo{author}{\bibfnamefont{A.~F.} \bibnamefont{Bangura}},
  \bibinfo{author}{\bibfnamefont{C.}~\bibnamefont{Kucharczyk}},
  \bibinfo{author}{\bibfnamefont{M.}~\bibnamefont{Johannes}}, \bibnamefont{and}
  \bibinfo{author}{\bibfnamefont{I.~R.} \bibnamefont{Fisher}},
  \bibinfo{journal}{arXiv:0902.1172}  (\bibinfo{year}{2009}).

\bibitem[{\citenamefont{Jia et~al.}(1995)\citenamefont{Jia, Callcott, Yurkas,
  Ellis, Himpsel, Samant, Stohr, Ederer, Carlisle, Hudson et~al.}}]{Callcott}
\bibinfo{author}{\bibfnamefont{J.~J.} \bibnamefont{Jia}},
  \bibinfo{author}{\bibfnamefont{T.~A.} \bibnamefont{Callcott}},
  \bibinfo{author}{\bibfnamefont{J.}~\bibnamefont{Yurkas}},
  \bibinfo{author}{\bibfnamefont{A.~W.} \bibnamefont{Ellis}},
  \bibinfo{author}{\bibfnamefont{F.~J.} \bibnamefont{Himpsel}},
  \bibinfo{author}{\bibfnamefont{M.~G.} \bibnamefont{Samant}},
  \bibinfo{author}{\bibfnamefont{J.}~\bibnamefont{Stohr}},
  \bibinfo{author}{\bibfnamefont{D.~L.} \bibnamefont{Ederer}},
  \bibinfo{author}{\bibfnamefont{J.~A.} \bibnamefont{Carlisle}},
  \bibinfo{author}{\bibfnamefont{E.~A.} \bibnamefont{Hudson}},
  \bibnamefont{et~al.}, \bibinfo{journal}{Rev. Sci. Instrum.}
  \textbf{\bibinfo{volume}{66}}, \bibinfo{pages}{1394} (\bibinfo{year}{1995}).

\bibitem[{\citenamefont{Thole and van~der Laan}(1988)}]{Thole}
\bibinfo{author}{\bibfnamefont{B.~T.} \bibnamefont{Thole}} \bibnamefont{and}
  \bibinfo{author}{\bibfnamefont{G.}~\bibnamefont{van~der Laan}},
  \bibinfo{journal}{Phys. Rev. B} \textbf{\bibinfo{volume}{38}},
  \bibinfo{pages}{3158} (\bibinfo{year}{1988}).

\bibitem[{\citenamefont{de~Groot et~al.}(1990)\citenamefont{de~Groot, Fuggle,
  Thole, and Sawatzky}}]{grootprb}
\bibinfo{author}{\bibfnamefont{F.~M.~F.} \bibnamefont{de~Groot}},
  \bibinfo{author}{\bibfnamefont{J.~C.} \bibnamefont{Fuggle}},
  \bibinfo{author}{\bibfnamefont{B.~T.} \bibnamefont{Thole}}, \bibnamefont{and}
  \bibinfo{author}{\bibfnamefont{G.~A.} \bibnamefont{Sawatzky}},
  \bibinfo{journal}{Phys. Rev. B} \textbf{\bibinfo{volume}{42}},
  \bibinfo{pages}{5459} (\bibinfo{year}{1990}).

\bibitem[{\citenamefont{van~der Laan and Kirkman}(1992)}]{Laan}
\bibinfo{author}{\bibfnamefont{G.}~\bibnamefont{van~der Laan}}
  \bibnamefont{and} \bibinfo{author}{\bibfnamefont{I.~W.}
  \bibnamefont{Kirkman}}, \bibinfo{journal}{J. Phys. Condens. Matter}
  \textbf{\bibinfo{volume}{4}}, \bibinfo{pages}{4189} (\bibinfo{year}{1992}).

\bibitem[{\citenamefont{Haule and Kotliar}(2008)}]{crystalfield}
\bibinfo{author}{\bibfnamefont{K.}~\bibnamefont{Haule}} \bibnamefont{and}
  \bibinfo{author}{\bibfnamefont{G.}~\bibnamefont{Kotliar}},
  \bibinfo{journal}{arXiv:0805.0722}  (\bibinfo{year}{2008}).

\bibitem[{\citenamefont{Moewes et~al.}(1998)\citenamefont{Moewes, Stadler,
  Winarski, Ederer, Grush, and Calcot}}]{CKdecay}
\bibinfo{author}{\bibfnamefont{A.}~\bibnamefont{Moewes}},
  \bibinfo{author}{\bibfnamefont{S.}~\bibnamefont{Stadler}},
  \bibinfo{author}{\bibfnamefont{R.~P.} \bibnamefont{Winarski}},
  \bibinfo{author}{\bibfnamefont{D.~L.} \bibnamefont{Ederer}},
  \bibinfo{author}{\bibfnamefont{M.~M.} \bibnamefont{Grush}}, \bibnamefont{and}
  \bibinfo{author}{\bibfnamefont{T.~A.} \bibnamefont{Calcot}},
  \bibinfo{journal}{Phys. Rev. B} \textbf{\bibinfo{volume}{58}},
  \bibinfo{pages}{15951} (\bibinfo{year}{1998}).

\bibitem[{\citenamefont{Chen et~al.}(1995)\citenamefont{Chen, Idzerda, Lin,
  Smith, Meigs, Chaban, Ho, Pellegrin, and Sette}}]{Femetal1}
\bibinfo{author}{\bibfnamefont{C.~T.} \bibnamefont{Chen}},
  \bibinfo{author}{\bibfnamefont{Y.~U.} \bibnamefont{Idzerda}},
  \bibinfo{author}{\bibfnamefont{H.~J.} \bibnamefont{Lin}},
  \bibinfo{author}{\bibfnamefont{N.~V.} \bibnamefont{Smith}},
  \bibinfo{author}{\bibfnamefont{G.}~\bibnamefont{Meigs}},
  \bibinfo{author}{\bibfnamefont{E.}~\bibnamefont{Chaban}},
  \bibinfo{author}{\bibfnamefont{G.~H.} \bibnamefont{Ho}},
  \bibinfo{author}{\bibfnamefont{E.}~\bibnamefont{Pellegrin}},
  \bibnamefont{and} \bibinfo{author}{\bibfnamefont{F.}~\bibnamefont{Sette}},
  \bibinfo{journal}{Phys. Rev. Lett.} \textbf{\bibinfo{volume}{75}},
  \bibinfo{pages}{152} (\bibinfo{year}{1995}).

\bibitem[{\citenamefont{Nakajima et~al.}(1999)\citenamefont{Nakajima, Stohr,
  and Idzerda}}]{Femetal2}
\bibinfo{author}{\bibfnamefont{R.}~\bibnamefont{Nakajima}},
  \bibinfo{author}{\bibfnamefont{J.}~\bibnamefont{Stohr}}, \bibnamefont{and}
  \bibinfo{author}{\bibfnamefont{Y.~U.} \bibnamefont{Idzerda}},
  \bibinfo{journal}{Phys. Rev. B} \textbf{\bibinfo{volume}{59}},
  \bibinfo{pages}{6421} (\bibinfo{year}{1999}).

\bibitem[{\citenamefont{Gao et~al.}(2006)\citenamefont{Gao, Qi, Tan, Wee, Yu,
  and Moser}}]{Femetal3}
\bibinfo{author}{\bibfnamefont{X.}~\bibnamefont{Gao}},
  \bibinfo{author}{\bibfnamefont{D.}~\bibnamefont{Qi}},
  \bibinfo{author}{\bibfnamefont{S.~C.} \bibnamefont{Tan}},
  \bibinfo{author}{\bibfnamefont{A.~T.~S.} \bibnamefont{Wee}},
  \bibinfo{author}{\bibfnamefont{X.}~\bibnamefont{Yu}}, \bibnamefont{and}
  \bibinfo{author}{\bibfnamefont{H.~O.} \bibnamefont{Moser}},
  \bibinfo{journal}{J. Electron Spec. Rel. Phenom.}
  \textbf{\bibinfo{volume}{151}}, \bibinfo{pages}{199} (\bibinfo{year}{2006}).

\bibitem[{\citenamefont{Duda et~al.}(2000)\citenamefont{Duda, Nordgren, Drager,
  Bocharov, and Kirchner}}]{DudaFe2O3}
\bibinfo{author}{\bibfnamefont{L.~C.} \bibnamefont{Duda}},
  \bibinfo{author}{\bibfnamefont{J.}~\bibnamefont{Nordgren}},
  \bibinfo{author}{\bibfnamefont{G.}~\bibnamefont{Drager}},
  \bibinfo{author}{\bibfnamefont{S.}~\bibnamefont{Bocharov}}, \bibnamefont{and}
  \bibinfo{author}{\bibfnamefont{T.}~\bibnamefont{Kirchner}},
  \bibinfo{journal}{J. Electron Spec. Rel. Phenom.}
  \textbf{\bibinfo{volume}{110}}, \bibinfo{pages}{257} (\bibinfo{year}{2000}).

\bibitem[{\citenamefont{Marusak et~al.}(1980)\citenamefont{Marusak, Messier,
  and White}}]{Fe2O3optic}
\bibinfo{author}{\bibfnamefont{L.~A.} \bibnamefont{Marusak}},
  \bibinfo{author}{\bibfnamefont{R.}~\bibnamefont{Messier}}, \bibnamefont{and}
  \bibinfo{author}{\bibfnamefont{W.~B.} \bibnamefont{White}},
  \bibinfo{journal}{J. Phys. Chem. Solids} \textbf{\bibinfo{volume}{41}},
  \bibinfo{pages}{981} (\bibinfo{year}{1980}).

\bibitem[{\citenamefont{Vernay et~al.}(2008)\citenamefont{Vernay, Moritz,
  Elfimov, Geck, Hawthorn, Devereaux., and Sawatzky}}]{Vernay}
\bibinfo{author}{\bibfnamefont{F.}~\bibnamefont{Vernay}},
  \bibinfo{author}{\bibfnamefont{B.}~\bibnamefont{Moritz}},
  \bibinfo{author}{\bibfnamefont{I.}~\bibnamefont{Elfimov}},
  \bibinfo{author}{\bibfnamefont{J.}~\bibnamefont{Geck}},
  \bibinfo{author}{\bibfnamefont{D.}~\bibnamefont{Hawthorn}},
  \bibinfo{author}{\bibfnamefont{T.~P.} \bibnamefont{Devereaux.}},
  \bibnamefont{and} \bibinfo{author}{\bibfnamefont{G.~A.}
  \bibnamefont{Sawatzky}}, \bibinfo{journal}{Phys. Rev. B}
  \textbf{\bibinfo{volume}{77}}, \bibinfo{pages}{104519}
  (\bibinfo{year}{2008}).

\bibitem[{\citenamefont{Meinders et~al.}(1993)\citenamefont{Meinders, Eskes,
  and Sawatzky}}]{Eskes}
\bibinfo{author}{\bibfnamefont{M.~B.~J.} \bibnamefont{Meinders}},
  \bibinfo{author}{\bibfnamefont{H.}~\bibnamefont{Eskes}}, \bibnamefont{and}
  \bibinfo{author}{\bibfnamefont{G.~A.} \bibnamefont{Sawatzky}},
  \bibinfo{journal}{Phys. Rev. B} \textbf{\bibinfo{volume}{48}},
  \bibinfo{pages}{3916} (\bibinfo{year}{1993}).

\bibitem[{\citenamefont{Kotani and Shin}(2001)}]{Kotani-Shin}
\bibinfo{author}{\bibfnamefont{A.}~\bibnamefont{Kotani}} \bibnamefont{and}
  \bibinfo{author}{\bibfnamefont{S.}~\bibnamefont{Shin}},
  \bibinfo{journal}{Rev. Mod. Phys.} \textbf{\bibinfo{volume}{73}},
  \bibinfo{pages}{203} (\bibinfo{year}{2001}).

\bibitem[{\citenamefont{Harrison}(2004)}]{Harrison}
\bibinfo{author}{\bibfnamefont{W.~A.} \bibnamefont{Harrison}},
  \emph{\bibinfo{title}{Elementary Electronic Structure}}
  (\bibinfo{publisher}{World Scientific}, \bibinfo{year}{2004}).

\bibitem[{\citenamefont{Dagotto et~al.}(2001)\citenamefont{Dagotto, Hotta, and
  Moreo}}]{Dagotto}
\bibinfo{author}{\bibfnamefont{E.}~\bibnamefont{Dagotto}},
  \bibinfo{author}{\bibfnamefont{T.}~\bibnamefont{Hotta}}, \bibnamefont{and}
  \bibinfo{author}{\bibfnamefont{A.}~\bibnamefont{Moreo}},
  \bibinfo{journal}{Physics Reports \textbf{344}, 1}  (\bibinfo{year}{2001}).

\bibitem[{\citenamefont{van~der Marel and Sawatzky}(1988)}]{Marel}
\bibinfo{author}{\bibfnamefont{D.}~\bibnamefont{van~der Marel}}
  \bibnamefont{and} \bibinfo{author}{\bibfnamefont{G.~A.}
  \bibnamefont{Sawatzky}}, \bibinfo{journal}{Phys. Rev. B}
  \textbf{\bibinfo{volume}{37}}, \bibinfo{pages}{10674} (\bibinfo{year}{1988}).

\bibitem[{\citenamefont{Mattheiss}(1964)}]{Mattheiss}
\bibinfo{author}{\bibfnamefont{L.}~\bibnamefont{Mattheiss}},
  \bibinfo{journal}{Phys. Rev.} \textbf{\bibinfo{volume}{133}},
  \bibinfo{pages}{A1399} (\bibinfo{year}{1964}).

\bibitem[{\citenamefont{Ankudinov and Rehr}(1997)}]{Ank97}
\bibinfo{author}{\bibfnamefont{A.~L.} \bibnamefont{Ankudinov}}
  \bibnamefont{and} \bibinfo{author}{\bibfnamefont{J.~J.} \bibnamefont{Rehr}},
  \bibinfo{journal}{Phys. Rev. B} \textbf{\bibinfo{volume}{56}},
  \bibinfo{pages}{R1712} (\bibinfo{year}{1997}).

\bibitem[{\citenamefont{von Barth and Hedin}(1972)}]{vonBarth}
\bibinfo{author}{\bibfnamefont{U.}~\bibnamefont{von Barth}} \bibnamefont{and}
  \bibinfo{author}{\bibfnamefont{L.}~\bibnamefont{Hedin}}, \bibinfo{journal}{J.
  Phys. C} \textbf{\bibinfo{volume}{5}}, \bibinfo{pages}{1629}
  (\bibinfo{year}{1972}).

\bibitem[{\citenamefont{Rehr and Albers}(1990)}]{rehralbers}
\bibinfo{author}{\bibfnamefont{J.~J.} \bibnamefont{Rehr}} \bibnamefont{and}
  \bibinfo{author}{\bibfnamefont{R.~C.} \bibnamefont{Albers}},
  \bibinfo{journal}{Phys. Rev. B} \textbf{\bibinfo{volume}{41}},
  \bibinfo{pages}{8139} (\bibinfo{year}{1990}).

\bibitem[{\citenamefont{Hedin and Lundqvist}(1969)}]{HL}
\bibinfo{author}{\bibfnamefont{L.}~\bibnamefont{Hedin}} \bibnamefont{and}
  \bibinfo{author}{\bibfnamefont{S.}~\bibnamefont{Lundqvist}},
  \bibinfo{journal}{Solid State Phys.} \textbf{\bibinfo{volume}{23}},
  \bibinfo{pages}{1} (\bibinfo{year}{1969}).

\bibitem[{\citenamefont{Lundqvist}(1967)}]{lundqvist}
\bibinfo{author}{\bibfnamefont{B.~I.} \bibnamefont{Lundqvist}},
  \bibinfo{journal}{Phys. Kondens. Mater.} \textbf{\bibinfo{volume}{6}},
  \bibinfo{pages}{192} (\bibinfo{year}{1967}).

\bibitem[{\citenamefont{de~Leon et~al.}(1991)\citenamefont{de~Leon, Rehr,
  Zabinsky, and Albers}}]{jose}
\bibinfo{author}{\bibfnamefont{J.~M.} \bibnamefont{de~Leon}},
  \bibinfo{author}{\bibfnamefont{J.~J.} \bibnamefont{Rehr}},
  \bibinfo{author}{\bibfnamefont{S.~I.} \bibnamefont{Zabinsky}},
  \bibnamefont{and} \bibinfo{author}{\bibfnamefont{R.~C.}
  \bibnamefont{Albers}}, \bibinfo{journal}{Phys. Rev. B}
  \textbf{\bibinfo{volume}{44}}, \bibinfo{pages}{4146} (\bibinfo{year}{1991}).

\bibitem[{\citenamefont{Blaha et~al.}(2001)\citenamefont{Blaha, Schwarz,
  Madsen, Kvasnicka, and Luitz}}]{wien}
\bibinfo{author}{\bibfnamefont{P.}~\bibnamefont{Blaha}},
  \bibinfo{author}{\bibfnamefont{K.}~\bibnamefont{Schwarz}},
  \bibinfo{author}{\bibfnamefont{G.}~\bibnamefont{Madsen}},
  \bibinfo{author}{\bibfnamefont{D.}~\bibnamefont{Kvasnicka}},
  \bibnamefont{and} \bibinfo{author}{\bibfnamefont{J.}~\bibnamefont{Luitz}},
  \emph{\bibinfo{title}{WIEN2k, An Augmented Plane Wave + Local Orbitals
  Program for Calculating Crystal Properties}} (\bibinfo{publisher}{Karlheinz
  Schwarz, Techn. Universität Wien, Austria}, \bibinfo{year}{2001}),
  \bibinfo{note}{iSBN 3-9501031-1-2}.

\bibitem[{\citenamefont{Johannes and Mazin}(2009)}]{Johannes}
\bibinfo{author}{\bibfnamefont{M.~D.} \bibnamefont{Johannes}} \bibnamefont{and}
  \bibinfo{author}{\bibfnamefont{I.}~\bibnamefont{Mazin}},
  \bibinfo{journal}{arXiv:0904.3857}  (\bibinfo{year}{2009}).

\bibitem[{\citenamefont{Kr\"{u}ger et~al.}(2009)\citenamefont{Kr\"{u}ger,
  Kumar, Zaanen, and van~den Brink}}]{Kruger09}
\bibinfo{author}{\bibfnamefont{F.}~\bibnamefont{Kr\"{u}ger}},
  \bibinfo{author}{\bibfnamefont{S.}~\bibnamefont{Kumar}},
  \bibinfo{author}{\bibfnamefont{J.}~\bibnamefont{Zaanen}}, \bibnamefont{and}
  \bibinfo{author}{\bibfnamefont{J.}~\bibnamefont{van~den Brink}},
  \bibinfo{journal}{Phys. Rev. B} \textbf{\bibinfo{volume}{79}},
  \bibinfo{pages}{054504} (\bibinfo{year}{2009}).

\bibitem[{\citenamefont{Hozoi and Fulde}(2009)}]{Hozoi09}
\bibinfo{author}{\bibfnamefont{L.}~\bibnamefont{Hozoi}} \bibnamefont{and}
  \bibinfo{author}{\bibfnamefont{P.}~\bibnamefont{Fulde}},
  \bibinfo{journal}{Physical Review Letters} \textbf{\bibinfo{volume}{102}},
  \bibinfo{pages}{136405} (\bibinfo{year}{2009}).

\bibitem[{\citenamefont{Singh}(2009)}]{Singh}
\bibinfo{author}{\bibfnamefont{R.}~\bibnamefont{Singh}},
  \bibinfo{journal}{arXiv:0903.4408}  (\bibinfo{year}{2009}).

\end{thebibliography}
\bibliographystyle{apsrev}
%\end{document}

\end{document}